 \documentclass[12pt]{article}
 \usepackage[cp1251]{inputenc} %- для WiN
 \usepackage[english, russian]{babel}
 \usepackage{amsmath, latexsym, amssymb, bm, array, graphics, eucal,
 amsfonts,}%mathscr
 \makeatletter
 \renewcommand{\@biblabel}[1]{#1.\hfill}

 \renewcommand{\Re}{\mathop{\rm Re}}
 \renewcommand{\Im}{\mathop{\rm Im\,}}
\renewcommand{\div}{\mathop{\rm div}}
\newcommand{\mc}[1]{\mathcal{#1}}
\newcommand{\E}{\mc{E}}
 \makeatother
 \renewcommand{\baselinestretch}{1.22}
\usepackage[dvips]{graphicx}
\usepackage[flushleft,small]{caption2} % чтобы в подписях к рисункам
\begin{document}
 \thispagestyle{empty}
 \large
 \renewcommand{\refname}{\begin{center} REFERENCES\end{center}}
\newcommand{\const}{\mathop{\rm const\, }}
 \begin{center}
\bf TRANSVERSAL ELECTRIC CONDUCTIVITY OF QUANTUM NON-DEGENERATE COLLISIONAL
PLASMAS
\end{center}\medskip
\begin{center}
  \bf A. V. Latyshev\footnote{$avlatyshev@mail.ru$},
  A. A. Yushkanov\footnote{$yushkanov@inbox.ru$}
\end{center}\medskip

\begin{center}
{\it Faculty of Physics and Mathematics,\\ Moscow State Regional
University, 105005,\\ Moscow, Radio str., 10--A}
\end{center}\medskip

\begin{abstract}
Formulas for calculation of transversal dielectric function and
transversal electric conductivity in quantum non-degenerate collisional
plasmas under
arbitrary degree of dege\-ne\-racy of the electron gas are received.
The Wigner--Vlasov--Boltzmann kinetic equation with collision
integral in BGK (Bhatnagar, Gross and Krook) form
in coordinate space is used. Various special cases are
in\-ves\-ti\-gated.  \medskip

\medskip

electron gas

\end{abstract}

\begin{center}\bf
INTRODUCTION
\end{center}

In the present work formulas for calculation of electric
conductivity and dielectric function in quantum non-degenerate collisional
plasma under arbitrary tempe\-ra\-tu\-re, i.e. under arbitrary degree of
degeneration of the electron gas are deduced.

During the derivation of the kinetic equation we generalize the
approach, developed by Klimontovich and Silin \cite {Klim}.

Dielectric function in the collisionless quantum gaseous plasma
was studied by many authors (see, for example, \cite {Klim} --
\cite{Shukla2}).

In the work \cite {Manf}, where the one-dimensional case
of the quantum plasma is investigated, the importance of derivation
of dielectric function with use of the quantum kinetic equation
with collision integral in the form of BGK -- model (Bhatnagar,
Gross, Krook) \cite{BGK}, \cite{Opher} was noted.

The present work is devoted to the performance of this task.% task.

A dielectric function is one of the most significant
characteristics of a plasma. This quantity is necessary for
description of the skin effect \cite{Gelder},
for analysis of surface plasmons \cite{Fuchs},
for description of the process of propagation and damping of the
transverse plasma oscillations \cite{Shukla2},
the mechanism of electromagnetic
waves penetration in plasma \cite{Shukla1}, and
for analysis of other problems of plasma physics \cite{Fuchs2},
\cite{Dressel}, \cite{Wier}, \cite{Brod} and \cite{Manf2}.

Kliewer and Fuchs were the first who have noticed \cite {Kliewer},
that the dielectric function for quantum plasma deduced by Lindhard
in collisional case does not pass into dielectric function for
classical plasma in the limit when Planck's constant $\hbar$
converges to zero. This means, that dielectric Lindhard's function
does not take into account electron collisions correctly. Kliewer
and Fuchs have corrected Lindhard's dielectric function
so that it passed into classical one under condition $ \hbar \to 0$.

In the works \cite{Fuchs}, \cite{Fuchs2} the dielectric function
received by them was applied to consideration of various questions
of metal optics.

In the work \cite{Mermin} the correct account of collisions in
framework of the relaxation model in electron momentum space
for the case of longi\-tu\-di\-nal dielectric function has been carried out. At
the same time the correct account of influence of collisions for
transversal dielectric function has not been implemented till
now.

The aim of the present work is the elimination of this lacuna. % specified
%above.

\begin{center}
  {\bf 1. KINETIC EQUATION FOR THE WIGNER FUNCTION}
\end{center}

Let's consider the Schr\"{o}dinger equation written for a particle
in an elect\-ro\-magnetic field in terms of density matrix $\rho$
$$
i\hbar \dfrac{\partial \rho}{\partial t}=H\rho-{H^*}'\rho.
\eqno{(1.1)}
$$

Here $H$ is the Hamilton operator, $H^*$ is the complex conjugate
operator to $H$, ${H^*}'$ is the complex conjugate operator to the
$H$, which forces on primed spatial variables $\mathbf{r}'$.

Hamilton operator for the free particle which is in the field of the
scalar potential $U$ and in the field of vector potential
$\mathbf{A}$, has the following form:
$$
H=\dfrac{(\mathbf{p}-\dfrac{e}{c}\mathbf{A})^2}{2m}+eU=$$$$=
\dfrac{\mathbf{p}^2}{2m}-
\dfrac{e}{2mc}(\mathbf{p}\mathbf{A}+\mathbf{A}\mathbf{p})
+\dfrac{e^2}{2mc^2}\mathbf{A}^2+eU.
\eqno{(1.2)}
$$

Here $\mathbf{p}$ is the momentum operator, $\mathbf{p}=-i\hbar \nabla$,
$e$ is the electron charge, $m$ is the electron mass, $c$ is the
light velocity.

Let's rewrite the Hamilton operator (1.2) in the explicit form:
$$
H=-\dfrac{\hbar^2}{2m}\triangle+\dfrac{ie\hbar}{2mc}\Big(2\mathbf{A}\nabla
+\nabla\mathbf{A}\Big)+
\dfrac{e^2}{2mc^2}\mathbf{A}^2+eU.
\eqno{(1.3)}
$$

Complex conjugate to the $H$ operator $H^*$ according to (1.3) has
the form
$$
H^*=-\dfrac{\hbar^2}{2m}\triangle-\dfrac{ie\hbar}{2mc}
\Big(2\mathbf{A}\nabla+\nabla\mathbf{A}\Big)+
\dfrac{e^2}{2mc^2}\mathbf{A}^2+eU.
$$

%The operators $H$ and ${H^*}'$, that were calculated in terms of
%density matrix have the following forms
Hence we can write down for $H\rho$:
$$
H\rho=-\dfrac{\hbar^2}{2m}\Delta \rho+\dfrac{ie\hbar}{2mc}
\Big(2\mathbf{A}\nabla \rho+\rho\nabla\mathbf{A}\Big)+
\dfrac{e^2}{2mc^2}\mathbf{A}^2\rho+eU \rho
\eqno{(1.4)}
$$
and for ${H^*}'\rho$:
$$
{H^*}'\rho=-\dfrac{\hbar^2}{2m}\Delta'\rho-\dfrac{ie\hbar}{2mc}
\Big(2\mathbf{A'}\nabla'\rho+\rho\nabla'\mathbf{A}\Big)+
\dfrac{e^2}{2mc^2}\mathbf{A'}^2\rho+eU' \rho.
\eqno{(1.5)}
$$

Operators $\nabla$ and $\Delta$ from Eqs (1.4) and (1.5) force on
unprimed spatial variables of the density matrix, i.e.
$\nabla=\nabla_{\mathbf{R}}$, $\Delta=\Delta_{\mathbf{R}}$.
In the operator ${H^*}'$ it is necessary to replace the operators
$\nabla=\nabla_{\mathbf{R}}$ and $\Delta=\Delta_{\mathbf{R}}$ by
operators $\nabla'\equiv\nabla_{\mathbf{R}'}$ and
$\Delta'\equiv\Delta_{\mathbf{R}'}$, in addition we introduce the
following designations
$$
\mathbf{A'}\equiv \mathbf{A}(\mathbf{R'},t),
\quad
\quad U'\equiv U(\mathbf{R'},t).
$$

Let's find the right-hand member of the equation (1.1), i.e.
difference between relations (1.4) and (1.5): $H\rho-{H^*}'\rho$.
According to (1.4) and (1.5) we have:
$$
H\rho-{H^*}'\rho=-\dfrac{\hbar}{2m}\Big(\Delta \rho-\Delta'\rho\Big)+
\hspace{8cm}
$$
$$
\hspace{1cm}+\dfrac{i e \hbar}{2mc}
\Big[2\Big(\mathbf{A}\nabla \rho+
\mathbf{A'}\nabla'\rho\Big)+\rho\Big(\nabla \mathbf{A}+
\nabla'\mathbf{A}\Big)\Big]+
$$
$$
\hspace{2cm}
+\dfrac{e^2}{2mc^2}\Big[\mathbf{A}^2(\mathbf{R},t)-\mathbf{A}^2(
\mathbf{R'},t)\Big]+
e[U(\mathbf{R},t)-U(\mathbf{R'},t)]\rho.
$$

The connection between density matrix
$\rho(\mathbf{r},\mathbf{r}',t)$ and Wigner function
\cite{Wigner}, \cite{Tatarskii}, \cite{Hillery}
$f(\mathbf{r},\mathbf{p},t)$ is
given by the inversion and direct Fourier conversions
$$
f(\mathbf{r},\mathbf{p},t)=\int
\rho(\mathbf{r}+\dfrac{\mathbf{a}}{2},\mathbf{r}-
\dfrac{\mathbf{a}}{2},t)e^{-i\mathbf{p}\mathbf{a}/\hbar}d^3a,
$$
$$
\rho(\mathbf{R},\mathbf{R}',t)=\dfrac{1}{(2\pi \hbar)^3}
\int f\Big(\dfrac{\mathbf{R}+\mathbf{R}'}{2}, \mathbf{p},t\Big)
e^{i\mathbf{p}(\mathbf{R}-\mathbf{R}')/\hbar}d^3p.
$$

The Wigner function is analogue of distribution function for quantum
systems. It is widely used in the diversified physics
questions.
Substituting the representation of the density matrix in terms
of the Wigner function (1.2) into the equation for the density matrix
(1.1), we obtain
$$
i\hbar\dfrac{\partial \rho}{\partial t}=H\left\{\dfrac{1}{(2\pi\hbar)^3}
\int f\Big(\dfrac{\mathbf{R}+\mathbf{R}'}{2},\mathbf{p'},t\Big)
e^{i\mathbf{p'}(\mathbf{R}-\mathbf{R}')/\hbar}\,d^3p'\right\}-
$$
$$
\hspace{1.8cm}-{H^*}'\left\{\dfrac{1}{(2\pi\hbar)^3}
\int f\Big(\dfrac{\mathbf{R}+\mathbf{R}'}{2},
\mathbf{p'},t\Big)
e^{i\mathbf{p'}(\mathbf{R}-\mathbf{R}')/\hbar}\,d^3p'\right\}.
$$

Let's use the equalities written above. Thus the right--hand member
of the previous equation we may present in explicit form. As a
result we receive the following equation:
$$
i\hbar \dfrac{\partial \rho}{\partial t}=
\dfrac{1}{(2\pi \hbar)^3}\int\left\{-\dfrac{i\hbar}{m}
\mathbf{p'}\nabla f +
\dfrac{ie\hbar}{2mc} \Big[\div{\mathbf{A}(\mathbf{R},t)}+
\div{\mathbf{A}(\mathbf{R}',t)}\Big]f+
\right.
$$
$$
+\left.\dfrac{ie\hbar}{2mc}\Big[\mathbf{A}(\mathbf{R},t)+
\mathbf{A}(\mathbf{R}',t)\Big]
\nabla f-\dfrac{e}{mc}\Big[\mathbf{A}(\mathbf{R},t)-
\mathbf{A}(\mathbf{R}',t)\Big]
\mathbf{p'}f\right.+
$$
$$
+\left.\dfrac{e^2}{2mc^2}\big[\mathbf{A}^2(\mathbf{R},t)-
\mathbf{A}^2(\mathbf{R}',t)\big]f+
e\big[U(\mathbf{R},t)-U(\mathbf{R}',t)\big]f\right\}
e^{i\mathbf{p'}(\mathbf{R}-\mathbf{R}')/\hbar}
d^3p'.
\eqno{(1.6)}
$$

In the equation (1.6) we will put
$$
\mathbf{R}=\mathbf{r}+\dfrac{\mathbf{a}}{2},\qquad
\mathbf{R}'=\mathbf{r}-\dfrac{\mathbf{a}}{2}.
$$

Then in this equation we obtain
$$
 f\Big(\dfrac{\mathbf{R}+\mathbf{R}'}{2}, \mathbf{p'},t\Big)
e^{i\mathbf{p'}(\mathbf{R}-\mathbf{R}')/\hbar}=f(\mathbf{r}, \mathbf{p'},t)
e^{i\mathbf{p'\,a}/\hbar}.
$$

Let's multiply the equation (1.6) by
$e^{-i\mathbf{p}\mathbf{a}/\hbar}$ and let's integrate it by
$\mathbf{a}$. Then we will divide both parts of the equation by
$i\hbar$. As a result we receive
$$
\dfrac{\partial f}{\partial t}=
\iint\Bigg\{-\dfrac{\mathbf{p'}}{m}\nabla f +
\dfrac{e}{2mc}
\Big[\mathbf{A}(\mathbf{r}+\dfrac{\mathbf{a}}{2},t)+
\mathbf{A}(\mathbf{r}-\dfrac{\mathbf{a}}{2},t)\Big]\nabla
f+ \medskip
$$
$$
+\dfrac{ie}{mc\hbar}\Big[
\mathbf{A}(\mathbf{r}+\dfrac{\mathbf{a}}{2},t)
-\mathbf{A}(\mathbf{r}-\dfrac{\mathbf{a}}{2},t)\Big]\mathbf{p'} f+
$$\medskip
$$+
\dfrac{e}{2mc}\Big[\div{\mathbf{A}(\mathbf{r}}+\dfrac{\mathbf{a}}{2},t)+
\div{\mathbf{A}(\mathbf{r}-\dfrac{\mathbf{a}}{2},t)}\Big]f-
$$
$$
-\dfrac{ie^2}{2mc^2\hbar}\Big[\mathbf{A}^2(\mathbf{r}+
\dfrac{\mathbf{a}}{2},t)-
\mathbf{A}^2(\mathbf{r}-\dfrac{\mathbf{a}}{2},t)\Big]f-
$$\medskip
$$
-\dfrac{ie}{\hbar}\Big[U(\mathbf{r}+\dfrac{\mathbf{a}}{2},t)-
U(\mathbf{r}-\dfrac{\mathbf{a}}{2},t)\Big]f\Bigg\}
e^{i(\mathbf{p'}-\mathbf{p})\mathbf{a}/\hbar}
\dfrac{d^3a\,d^3p'}{(2\pi\hbar)^3}.
\eqno{(1.7)}
$$

On the left-hand side of the equation (1.7) we have
$f=f(\mathbf{r},\mathbf{p},t)$, in the integral we have
$f=f(\mathbf{r},\mathbf {p'},t)$.

We consider the integral
$$
\iint\mathbf{p'}(\nabla f)e^{i(\mathbf{p'}-\mathbf{p})\mathbf{a}/\hbar}
\dfrac{d^3a\,d^3p'}{(2\pi\hbar)^3}=
\nabla\iint\mathbf{p'} f e^{i(\mathbf{p'}-\mathbf{p})\mathbf{a}/\hbar}
\dfrac{d^3a\,d^3p'}{(2\pi\hbar)^3}=
$$
$$
=\nabla\int\mathbf{p'}f
\delta(\mathbf{p'}-\mathbf{p})d\,\mathbf{p'}=\mathbf{p}\nabla
f(\mathbf{r},\mathbf{p}).
$$

Two following equalities can be verified similarly:
$$
\iint\dfrac{e}{mc}\mathbf{A}(\mathbf{r},t)(\nabla f(\mathbf{r},
\mathbf{p}',t))
e^{i(\mathbf{p'}-\mathbf{p})\mathbf{a}/\hbar}
\dfrac{d^3a\,d^3p'}{(2\pi\hbar)^3}=$$$$=
\dfrac{e}{mc}\mathbf{A}(\mathbf{r},t) \nabla f(\mathbf{r},\mathbf{p},t),
$$
and
$$
\iint\dfrac{e}{mc}(\div{\mathbf{A}(\mathbf{r},t)}) f(\mathbf{r},
\mathbf{p}',t)
e^{i(\mathbf{p'}-\mathbf{p})\mathbf{a}/\hbar}
\dfrac{d^3a\,d^3p'}{(2\pi\hbar)^3}=$$$$=
\dfrac{e}{mc}(\div{\mathbf{A}}(\mathbf{r},t)) f(\mathbf{r},\mathbf{p},t).
$$

Then the equation (1.6) can be rewritten as following
$$
\dfrac{\partial f}{\partial t}+\dfrac{1}{m}
\Big(\mathbf{p}-\dfrac{e}{c}\mathbf{A}\Big)\nabla f-\dfrac{e}{mc}
(\div{\mathbf{A}(\mathbf{r},t)})f(\mathbf{r},\mathbf{p},t)=
W[f].
\eqno{(1.8)}
$$

In the equation (1.8) the symbol $W[f]$ is the Wigner
--- Vlasov integral, defined by the equality
$$
W[f]=
\iint\left\{
\dfrac{e}{2mc}
\Big[\mathbf{A}(\mathbf{r}+\dfrac{\mathbf{a}}{2},t)+
\mathbf{A}(\mathbf{r}-\dfrac{\mathbf{a}}{2},t)-
2\mathbf{A}(\mathbf{r},t)\Big]\nabla f\right.+
$$\medskip
$$
+\dfrac{ie}{ mc\hbar}\Big[
\mathbf{A}(\mathbf{r}+\dfrac{\mathbf{a}}{2},t)
-\mathbf{A}(\mathbf{r}-\dfrac{\mathbf{a}}{2},t)\Big]\mathbf{p'}
f+$$$$+\dfrac{e}{2mc}\Big[\div{\mathbf{A}(\mathbf{r}+
\dfrac{\mathbf{a}}{2},t)
}+\div{\mathbf{A}(\mathbf{r}-\dfrac{\mathbf{a}}{2},t)}-
2\div{\mathbf{A}(\mathbf{r},t)}\Big]f-
$$\medskip
$$-
\dfrac{i e^2}{2 mc^2\hbar}\Big[\mathbf{A}^2(\mathbf{r}+
\dfrac{\mathbf{a}}{2},t)-
\mathbf{A}^2(\mathbf{r}-\dfrac{\mathbf{a}}{2},t)\Big]f-
$$\medskip
$$
-\left. \dfrac{ie}{\hbar}\Big[U(\mathbf{r}+\dfrac{\mathbf{a}}{2},t)-
U(\mathbf{r}-\dfrac{\mathbf{a}}{2},t)\Big]f\right\}
e^{i(\mathbf{p'}-\mathbf{p})\mathbf{a}/\hbar}
\dfrac{d^3a\,d^3p'}{(2\pi\hbar)^3}.
\eqno{(1.9)}
$$

The energy of the particle is equal to
$$
\E=\dfrac{1}{2m}\Big(\mathbf{p}-
\dfrac{e}{c}\mathbf{A}\Big)^2+eU.
$$

Then the velocity of the particle $\mathbf{v}$ is equal to
$$
\mathbf{v}=\dfrac{\partial \E}{\partial \mathbf{p}}=
\dfrac{1}{m}\Big(\mathbf{p}-\dfrac{e}{c}\mathbf{A}\Big),
$$
besides,
$$
\nabla \mathbf{v}=-\dfrac{e}{mc}\div{\mathbf{A}}.
$$

Hence, the left--hand part of the equation (1.9) equals to:
$$
\dfrac{\partial f}{\partial t}+\dfrac{1}{m}
\Big(\mathbf{p}-\dfrac{e}{c}\mathbf{A}\Big)\nabla f-f\dfrac{e}{mc}
\div{\mathbf{A}}=
\dfrac{\partial f}{\partial t}+\mathbf{v}\nabla f+f\nabla \mathbf{A}=
$$
$$
=\dfrac{\partial f}{\partial t}+\nabla(\mathbf{v}f).
$$

Therefore the equation (1.9) can be rewritten in standard for
transport theory form
$$
\dfrac{\partial f}{\partial t}+\nabla (\mathbf{v}f)=W[f].
\eqno{(1.10)}
$$

In the case of collisional plasma we may write the kinetic equation
(1.10) as following
$$
\dfrac{\partial f}{\partial t}+\nabla (\mathbf{v}f)=B[f,f]+
W[f].
\eqno{(1.11)}
$$

In the equation (1.11) the symbol $B[f,f]$ represents the collision
integral.

\begin{center}\bf
2. RELAXATION MODEL OF KINRTIC EQUATION
\end{center}

Under the electron scattering on impurity we will consider the
equation (1.11) with collision integral in the form of relaxation
$\tau$--model \cite{BGK}, \cite{Opher}:
$$
\dfrac{\partial f}{\partial t}+\nabla(\mathbf{v}f)=
\dfrac{f^{(0)}-f}{\tau}+W[f].
\eqno{(2.1)}
$$

In the equation (2.1) $\tau$ is the mean time between two
consecutive collisions, $\tau=1/\nu$, $\nu$ is the collision
frequency, $f^{(0)}$ is the local equilibrium Fermi --- Dirac
distribution function,
$$
f^{(0)}=\Big[1+\exp\Big(\dfrac{\E-\mu}{k_BT}\Big)\Big]^{-1}.
$$

Here $k_B$ is the Boltzmann constant, $T$ is the plasma temperature,
$\E$ is the electron energy,
$\mu$ is the chemical potential of electron gas.

In an explicit form the local equilibrium distribution function has
the following form
$$
f^{(0)}(\mathbf{r},t)=\Bigg[1+\exp\Big[\dfrac{\big[\mathbf{p}-
(e/c)\mathbf{A}(\mathbf{r},t)\big]^2}{2mk_BT}+
\dfrac{eU(\mathbf{r},t)-\mu}{k_BT}\Big]\Bigg]^{-1}.
$$

We introduce the dimensionless electron velocity
$\mathbf{C}(\mathbf{r},t)$, scalar potential $\phi(\mathbf{r},t)$
and chemical potential $\alpha$:
$$
\mathbf{C}(\mathbf{r},t)=\dfrac{\mathbf{v}(\mathbf{r},t)}{v_T},\qquad
\phi(\mathbf{r},t)=\dfrac{eU(\mathbf{r},t)}{k_BT}, \qquad
\alpha=\dfrac{\mu}{k_BT},
$$
where
$v_T=\dfrac{1}{\sqrt{\beta}}$ is the thermal electron velocity,
$\beta=\dfrac{m}{2k_BT}$.

Now local equilibrium function can be presented in terms of the
electron velocity as follows
$$
f^{(0)}(\mathbf{r},t)=\Big[1+\exp\Big(\dfrac{mv^2(\mathbf{r},t)}{2k_BT}+
\dfrac{eU(\mathbf{r},t)-\mu}{k_BT}\Big)
\Big]^{-1},
$$
or, in dimensionless parameters,
$$
f^{(0)}(\mathbf{r},t)=\dfrac{1}{1+\exp\big[C^2(\mathbf{r},t)+
\phi(\mathbf{r},t)-\alpha\big]}.
\eqno{(2.2)}
$$

We designate \(\chi=\alpha-\phi\). Then we have $
f^{(0)}=\dfrac{1}{1+e^{C^2-\chi}}. $

The quantity $\chi $ is defined from the conservation law
of number of particles
$$
\int f d\Omega_F=
\int f^{(0)}d\Omega_F.
$$

Here $d\Omega_F$ is the quantum measure for electrons,
$$
d\Omega_F=\dfrac{2d^3p}{(2\pi\hbar)^3}.
$$

Let's note, that in the case of constant potentials $U={\rm const},
\mathbf{A}=\const$ the equilibrium distribution function (2.2) is
the solution of the equation (2.1).

Let's find the electron concentration (numerical density) $N$ and
mean electron velocity $\mathbf{u}$ in an equilibrium state. These
macroparameters are defined as follows:
$$
N(\mathbf{r},t)=\int f(\mathbf{r},\mathbf{p},t)d\Omega_F,
$$
$$
\mathbf{u}(\mathbf{r},t)=\dfrac{1}{N(\mathbf{r},t)}
\int \mathbf{v}(\mathbf{r},t)
f(\mathbf{r},\mathbf{p},t)d\Omega_F.
$$

For calculation of these macroparameters in equilibrium condition it
is necessary to put $f=f^{(0)}$, where $f^{(0)}$ is defined by
equality (1.13). We designate these macroparameters in equilibrium
condition through $N^{(0)}(\mathbf{r},t)$ and
$\mathbf{u}^{(0)}(\mathbf{r},t)$.

Let's carry out the replacement of the integration variable
$\mathbf{p}-(e/c)\mathbf{A}(\mathbf{r},t)=\mathbf{p}'$ in the
previous equalities. Then, passing to integration in spherical
co\-or\-di\-na\-tes, for numerical density in an equilibrium state
we get:
$$
N^{(0)}=\dfrac{m^3v_T^3}{\pi^2\hbar^3}f_2(\alpha-\phi),
\eqno{(2.3)}
$$
where
$$
f_2(\alpha-\phi)=\int\limits_{0}^{\infty}\dfrac{x^2\;dx}{1+\exp(x^2+
\phi-\alpha)}=\int\limits_{0}^{\infty}x^2f_F(\alpha-\phi)\,dx.
$$

In the same way, as for numerical density, for mean velocity in
equilibrium state we derive
$$
\mathbf{u}^{(0)}(\mathbf{r},t)=\dfrac{1}{N^{(0)}}
\int \mathbf{v}(\mathbf{r},t)f^{(0)}(\mathbf{r},\mathbf{p},t)d\Omega_F,
%\eqno{(1.15)}
$$
or, in explicit form,
$$
\mathbf{u}^{(0)}(\mathbf{r},t)=\dfrac{2}{N^{(0)}(2\pi\hbar)^3}\int
\dfrac{[\mathbf{p}-(e/c)\mathbf{A}]\;d^3p}
{1+\exp\Big[\dfrac{(\mathbf{p}-(e/c)\mathbf{A})^2}{2k_BTm}+
\dfrac{eU-\mu}{k_Tm}\Big]}.
$$

After the same change of variables
$\mathbf{p}-(e/c)\mathbf{A}(\mathbf{r},t)= \mathbf{p}'$ we receive:
$$
\mathbf{u}^{(0)}(\mathbf{r},t)=
\dfrac{2}{N^{(0)}(2\pi\hbar)^3}\int \dfrac{\mathbf{p}'\;d^3p'}
{1+\exp\Big[\dfrac{{p'}^2}{2k_BTm}+\dfrac{eU-\mu}{k_Tm}\Big]}=0.
\eqno{(2.4)}
$$

So, the electron velocity in an equilibrium state according to
$(2.4)$ is equal to zero.

Let's note, that numerical electron density and their mean velocity
satisfy the usual continuity equation:
$$
\dfrac{\partial N}{\partial t}+\mathbf{div}(N\mathbf{u})=0.
\eqno{(2.5)}
$$

For the derivation of the continuity equation (2.5) it is necessary
to integrate the kinetic equation (2.1) by quantum measure for
electrons $d\Omega_F$ and to use the definition of numerical density
and mean velocity.
Then it is necessary to use the conservation law of number of
particles and to check up, if the integral by quantum measure
$d\Omega_F$ of Wigner --- Vlasov integral is equal to zero. Indeed,
we have
$$
\int W[f]\dfrac{2\;d^3p}{(2\pi\hbar)^3}=2\iint
\Big\{\cdots\Big\}
e^{i\mathbf{p'}\mathbf{a}/\hbar}\delta(\mathbf{a})\,d^3a\,d^3p'=
$$
$$
=2\int \Big\{\cdots\Big\}\Bigg|_{\mathbf{a}=0}d^3p\equiv 0,
$$
as after some algebra,
$$
 \Big\{\cdots\Big\}\Bigg|_{\mathbf{a}=0}\equiv 0.
$$

Here the symbol $\{\cdots\}$ means the same expression, as in the
right-hand side of the equation (1.9).

Let's note, that the left-hand side of the kinetic equation (1.11)
or (2.1) takes standard form for transport theory under the
following gauge condition:
$$
\div{\mathbf{A}(\mathbf{r},t)}=0.
\eqno{(2.6)}
$$

Thus, i.e. in case of gauge (2.6), the kinetic equation has
the following form:
$$
\dfrac{\partial f}{\partial t}+\mathbf{v}\nabla f=B[f,f]+
W[f].
\eqno{(2.7)}
$$

Here the Wigner --- Vlasov integral equals to:
$$
W[f]=
\iint\left\{
\dfrac{e}{2mc}
\Big[\mathbf{A}(\mathbf{r}+\dfrac{\mathbf{a}}{2},t)+
\mathbf{A}(\mathbf{r}-\dfrac{\mathbf{a}}{2},t)-
2\mathbf{A}(\mathbf{r},t)\Big]\nabla f\right.+
$$
\hspace{0.5cm}
$$
+\dfrac{ie}{ mc\hbar}\Big[
\mathbf{A}(\mathbf{r}+\dfrac{\mathbf{a}}{2},t)
-\mathbf{A}(\mathbf{r}-\dfrac{\mathbf{a}}{2},t)\Big]\mathbf{p'}
f-\dfrac{i e^2}{2 mc^2\hbar}
\Big[\mathbf{A}^2(\mathbf{r}+\dfrac{\mathbf{a}}{2},t)-
\mathbf{A}^2(\mathbf{r}-\dfrac{\mathbf{a}}{2},t)\Big]f-
$$
\hspace{0.5cm}
$$
-\left. \dfrac{ie}{\hbar}\Big[U(\mathbf{r}+\dfrac{\mathbf{a}}{2},t)-
U(\mathbf{r}-\dfrac{\mathbf{a}}{2},t)\Big]f\right\}
e^{i(\mathbf{p'}-\mathbf{p})\mathbf{a}/\hbar}
\dfrac{d^3a\,d^3p'}{(2\pi\hbar)^3}.
\eqno{(2.8)}
$$\\

\begin{center}
  {\bf 3. LINEARIZATION OF THE KINETIC EQUATION AND ITS SOLUTION}
\end{center}

Let's consider the kinetic equation with collision integral in the
form of $\tau$--model and suppose, that the scalar potential is
equal to zero: $U(\mathbf {r},t)\equiv 0$.

We take vector potential which is orthogonal to the direction of the
wave vector $\mathbf {k}:\mathbf{k}\mathbf{A}=0$ in the form of
a running harmonic wave:
$$
\mathbf{A}(\mathbf{r},t)=\mathbf{A}_0
e^{i(\mathbf{k}\mathbf{r}-\omega t)}.
$$

%The gauge condition is satisfied automatically under
%such configuration of the vector potential (1.22).
%In addition, the first term in the right-hand side
%of the Wigher -- Vlasov integral equals to zero.

We suppose that the vector potential is small enough. This
assump\-tion allows us to linearize the equation and to neglect terms
quadratic in electric field.

Then the equation (2.7) can be reduced to:
$$
\dfrac{\partial f}{\partial t}+\mathbf{v}\nabla f=
\dfrac{f^{(0)}-f}{\tau}+W[f].
\eqno{(3.1)}
$$

In this case chemical potential is equal to a constant.

In the equation (3.1) local equilibrium Fermi --- Dirac distribution
is simplified as following:
$$
f^{(0)}=\Big[1+\exp\Big(C^2(\mathbf{r},t)-\alpha\Big)\Big]^{-1}.
\eqno{(3.2)}
$$

The Wigner --- Vlasov integral (2.8) also can be simplified
essentially and has the following form:
$$
W[f]=\dfrac{ie}{mc\hbar }
\iint
\Big[\mathbf{A}(\mathbf{r}+\dfrac{\mathbf{a}}{2},t)
-\mathbf{A}(\mathbf{r}-\dfrac{\mathbf{a}}{2},t)\Big]\mathbf{p'}
f^{}
e^{i(\mathbf{p'}-\mathbf{p})\mathbf{a}/\hbar}
\dfrac{d^3a\,d^3p'}{(2\pi\hbar)^3}.
\eqno{(3.3)}
$$

We notice, that
$$
\mathbf{A}(\mathbf{r}+\dfrac{\mathbf{a}}{2},t)
-\mathbf{A}(\mathbf{r}-\dfrac{\mathbf{a}}{2},t)=\mathbf{A}(\mathbf{r},t)
\Big[e^{i\mathbf{k}\mathbf{a}/2}-e^{-i\mathbf{k}\mathbf{a}/2}\Big].
$$

Calculating the integral in (3.3), we find, that
$$
W[\mathbf{A},f^{}]=\dfrac{ie}{mc\hbar}\mathbf{A}(\mathbf{r},t)
\iint \Big[e^{i\mathbf{k}\mathbf{a}/2}-e^{-i\mathbf{k}\mathbf{a}/2}\Big]
e^{i(\mathbf{p}'-\mathbf{p})\mathbf{a}/\hbar}\dfrac{d^3a\,d^3p'}
{(2\pi\hbar)^3}.
$$

The internal integral is equal to:
$$
\dfrac{1}{(2\pi\hbar)^3}\int \Big\{\exp\Big(i\Big[\mathbf{p}'-\mathbf{p}+
\frac{\mathbf{k}\hbar}{2}\Big]\dfrac{\mathbf{a}}{\hbar}\Big)-
\exp\Big(i\Big[\mathbf{p}'-\mathbf{p}+
\frac{\mathbf{k}\hbar}{2}\Big]\dfrac{\mathbf{a}}{\hbar}\Big)\Big\}d^3a=
$$
$$
=\delta\Big(\mathbf{p}'-\mathbf{p}+\dfrac{\hbar\mathbf{k}}{2}\Big)-
\delta\Big(\mathbf{p}'-\mathbf{p}-\dfrac{\hbar\mathbf{k}}{2}\Big).
$$

We calculate the Wigner --- Vlasov integral
$$
W[f,\mathbf{A}]=
$$
%\vspace{0.1cm}
$$=\mathbf{A}(\mathbf{r},t)\dfrac{ie}{mc\hbar}
\int \Big[\delta(\mathbf{p}'-\mathbf{p}+\dfrac{\hbar\mathbf{k}}{2})-
\delta(\mathbf{p}'-\mathbf{p}-\dfrac{\hbar\mathbf{k}}{2})\Big]
\mathbf{p}'f(\mathbf{r},\mathbf{p}',t)\,d^3p'=
$$\vspace{0.3cm}
$$
=\mathbf{A}(\mathbf{r},t)\dfrac{ie}{mc\hbar}
\Big[\Big(\mathbf{p}-\dfrac{\hbar \mathbf{k}}{2}\Big)f(\mathbf{r},
\mathbf{p}-\dfrac{\hbar \mathbf{k}}{2},t)-
\Big(\mathbf{p}+\dfrac{\hbar \mathbf{k}}{2}\Big)f(\mathbf{r},
\mathbf{p}+\dfrac{\hbar \mathbf{k}}{2},t)\Big]=
$$\vspace{0.3cm}
$$
=\mathbf{A}(\mathbf{r},t)\dfrac{ie}{mc\hbar}
\Big\{\mathbf{p}\Big[f(\mathbf{r},
\mathbf{p}-\dfrac{\hbar \mathbf{k}}{2},t)-f(\mathbf{r},
\mathbf{p}+\dfrac{\hbar \mathbf{k}}{2},t)\Big]-
$$
\vspace{0.3cm}
$$
-\dfrac{\hbar\mathbf{k}}{2}\Big[f(\mathbf{r},
\mathbf{p}-\dfrac{\hbar \mathbf{k}}{2},t)+f(\mathbf{r},
\mathbf{p}+\dfrac{\hbar \mathbf{k}}{2},t)\Big]\Big\}=
$$\vspace{0.3cm}
$$
=\mathbf{A}(\mathbf{r},t)\dfrac{ie}{mc\hbar}\mathbf{p}\Big(f_+-f_-\Big),
$$
where
$$
f_{\pm}\equiv f(\mathbf{r},\mathbf{p}\mp\dfrac{\hbar
\mathbf{k}}{2},t).
$$

Consequently, the Wigner --- Vlasov integral is equal to
$$
W[f]=\dfrac{ie}{mc\hbar}\mathbf{p}\mathbf{A}
(\mathbf{r},t)\Big[f_+^{}-f_-^{}\Big]=
\dfrac{iep_T}{mc\hbar}\mathbf{P}\mathbf{A}(\mathbf{r},t)
\Big[f_+^{}-f_-^{}\Big]=$$$$=\dfrac{iev_T}{c\hbar}\mathbf{P}\mathbf{A}
(\mathbf{r},t)\Big[f_+^{}-f_-^{}\Big].
\eqno{(3.4)}
$$

Here and below the expression $\mathbf{P}\mathbf{A}$ means
scalar production.

Further we will use dimensionless velocity $\mathbf{C}$ in the form
$$
\mathbf{C}=\dfrac{\mathbf{v}}{v_T}=\dfrac{\mathbf{p}}{p_T}-
\dfrac{e}{cp_T}\mathbf{A}(\mathbf{r},t)\equiv \mathbf{P}-
\dfrac{e}{cp_T}\mathbf{A}(\mathbf{r},t),
$$
where $\mathbf{P}=\dfrac{\mathbf{p}}{p_T}$ is the dimensionless momentum.

In linear approximation it is possible to replace the function $f$
in Wigner --- Vlasov integral by the absolute Fermi --- Dirac
distribution, i.e. we put $f=f_F(P)$, where
$$
f_F(P)=\dfrac{1}{1+\exp (P^2-\alpha)}, \qquad \alpha=\const.
$$

Here Wigner --- Vlasov integral (3.4) has the following form:
$$
W[f_F]=\dfrac{iev_T}{c\hbar}\mathbf{P}\mathbf{A}
(\mathbf{r},t)\Big[f_F^+-f_F^-\Big],
$$
where
$$
f_F^{\pm}\equiv f_F^{\pm}(\mathbf{P})=
\dfrac{1}{1+
\exp\Big[\Big(\mathbf{P}\mp \dfrac{\hbar \mathbf{k}}{2p_T}\Big)^2
-\alpha\Big]},
$$
and $p_T=mv_T$ is the thermal electron momentum, or,
$$
f_F^{\pm}=\dfrac{1}{1+e^{P^2_{\pm}-\alpha}}.
$$

Here
$$
P^2_{\pm}=\Big(\mathbf{P}\mp\dfrac{\hbar \mathbf{k}}{2p_T}\Big)^2
=\Big(P_x\mp\dfrac{\hbar k_x}{2p_T}\Big)^2+
\Big(P_y\mp\dfrac{\hbar k_y}{2p_T}\Big)^2+
\Big(P_z\mp\dfrac{\hbar k_z}{2p_T}\Big)^2,
$$
or
$$
P^2_{\pm}=\dfrac{\Big(p_x\mp\dfrac{\hbar k_x}{2}\Big)^2+
\Big(p_y\mp\dfrac{\hbar k_y}{2}\Big)^2+
\Big(p_z\mp\dfrac{\hbar k_z}{2}\Big)^2}{p_T^2}.
$$

The linearization of the Wigner equilibrium function (3.2) we will
carry out in terms of vector potential $\mathbf{A}(\mathbf{r},t)$:
$$
f^{(0)}=f^{(0)}\Big|_{\mathbf{A}=0}+
\dfrac{\partial f^{(0)}}
{\partial \mathbf{A}}\Bigg|_{\mathbf{A}=0}%\right.
 \mathbf{A}(\mathbf{r},t),
$$
or, in explicit form:
$$
f^{(0)}=f_F(P)+g(P)\dfrac{2e}{cp_T}\mathbf{P}\mathbf{A}(\mathbf{r},t),
\eqno{(3.5)}
$$
$$
g(P)=\dfrac{e^{P^2-\alpha}}{(1+e^{P^2-\alpha})^2}.
$$

Considering decomposition (3.5), we will search for Wigner's function
in the form:
$$
f=f_F(P)+g(P)\dfrac{2e}{cp_T}\mathbf{P}\mathbf{A}(\mathbf{r},t)+
g(P)(\mathbf{P} \mathbf{A}(\mathbf{r},t))h(\mathbf{P}).%+g(P)\delta \alpha.
\eqno{(3.6)}
$$

We receive the following equation
$$
 \mathbf{P}\mathbf{A}(\mathbf{r},t)\;g(P)(\nu-i\omega+
i\mathbf{k} \mathbf{v})h(\mathbf{P})=
$$
$$
=\mathbf{P}\mathbf{A}\Bigg[\dfrac{2ie}{cp_T}g(P)(\omega-v_T\mathbf{k} \mathbf{P})+\dfrac{iev_T}{c\hbar}
(\mathbf{r},t)(f_F^+-f_F^-)\Bigg].
$$

From this equation we find
$$
 \mathbf{P}\mathbf{A}(\mathbf{r},t)g(P)h(\mathbf{P})=
\mathbf{P} \mathbf{A}(\mathbf{r},t)
\dfrac{2ie}{cp_T}\dfrac{\omega-v_T \mathbf{k} \mathbf{P}}
{\nu-i\omega+iv_T\mathbf{k} \mathbf{P}}g(P)+
$$
$$
+\dfrac{iev_T}{c\hbar}\mathbf{P} \mathbf{A}(\mathbf{r},t)
\dfrac{f_F^+-f_F^-}{\nu-i\omega+iv_T\mathbf{k} \mathbf{P}}.
\eqno{(3.7)}
$$

With the help of (3.6) and (3.7) we construct the full distribution
function
$$
f=f^{(0)}+g(P)\mathbf{P A}h(\mathbf{P})=
$$
$$
=f^{(0)}+\dfrac{2ie}{cp_T}\dfrac{\omega-v_T\mathbf{k P}}
{\nu-i \omega +iv_T\mathbf{k P}}g(P)\mathbf{P A}+
\dfrac{iev_T}{c\hbar}\dfrac{\mathbf{P A}(f_F^+-f_F^-)}
{\nu-i \omega +iv_T\mathbf{k P}},
%\eqno{(2.8)}
$$ \medskip
or
$$
f=f^{(0)}+\mathbf{P A}\Bigg[\dfrac{2ie}{cp_T}
\dfrac{\omega\tau-\mathbf{k}_1\mathbf{P}}
{1-i \omega\tau +i\mathbf{k}_1\mathbf{P}}g(P)
+\dfrac{iel}{c\hbar}\dfrac{f_F^+-f_F^-}{1-i \omega\tau +i\mathbf{k}_1\mathbf{P}}
\Bigg].
\eqno{(3.8)}
$$ \medskip

Here $\mathbf{k}_1=\mathbf{k}l$, $l$ is the electron mean free path,
$\;l=v_T\tau$, $\mathbf{k}_1$ is the dimensionless wave vector.

\begin{center}
\bf 4. DENSITY OF ELECTRIC CURRENT
\end{center}

We consider the connection between electric field and potentials
$$
\mathbf{E}(\mathbf{r},t)=
-\dfrac{1}{c}\dfrac{\partial \mathbf{A}(\mathbf{r},t)}{\partial t}-
\dfrac{\partial U(\mathbf{r},t)}{\partial \mathbf{r}},
$$
or
$$
\mathbf{E}(\mathbf{r},t)=\dfrac{i \omega}{c}\mathbf{A}(\mathbf{r},t).
$$

Hence, the current density is connected with vector potential as:
$$
\mathbf{j}(\mathbf{r},t)=
\sigma_{tr}\dfrac{i \omega}{c}\mathbf{A}(\mathbf{r},t).
$$

By definition, the current density is equal to
$$
\mathbf{j}(\mathbf{r},t)=e\int \mathbf{v}f\dfrac{2\,d^3p}{(2\pi\hbar)^3}.
$$

Let's note, that the current density in the equilibrium state is equal to zero:
$$
\mathbf{j}^{(0)}(\mathbf{r},t)=
e\int \mathbf{v}(\mathbf{r},\mathbf{v},t)f^{(0)}
\dfrac{2p_T^3\,d^3P}{(2\pi\hbar)^3}=0.
$$

Indeed, considering that mean electron velocity in the equilibrium
state is equal to zero, according to $(2.4)$ we have:
$$
\mathbf{j}^{(0)}(\mathbf{r},t)=
e N^{(0)}\mathbf{u}^{(0)}(\mathbf{r},t)\equiv 0.
$$

Hence, with the use of equality (3.8) we have the following equality:
$$
\mathbf{j}(\mathbf{r},t)=\dfrac{2ep_T^3}{(2\pi\hbar)^3}
\int \big(\mathbf{PA}\big)\mathbf{v}(\mathbf{r},\mathbf{v},t)\times $$$$ \times\Bigg[
\dfrac{2ie}{cp_T}\dfrac{\omega\tau-\mathbf{k}_1\mathbf{P}}
{1-i \omega\tau +i\mathbf{k}_1 \mathbf{P}}\,g(P)+%$$$$+
\dfrac{iel}{c\hbar}\dfrac{f_F^+-f_F^-}
{1-i \omega\tau +i\mathbf{k}_1 \mathbf{P}}\Bigg]d^3P.
$$

Substituting obvious expression for the velocity into this equality
$$
\mathbf{v}(\mathbf{r},t)=
\dfrac{\mathbf{p}}{m}-\dfrac{e \mathbf{A}(\mathbf{r},t)}{mc}=
\dfrac{p_T\mathbf{P}}{m}-\dfrac{e \mathbf{A}(\mathbf{r},t)}{mc},
$$
and, after linearization of it by vector field, we receive
$$
\mathbf{j}(\mathbf{r},t)=
\dfrac{2ep_T^4}{(2\pi \hbar)^3m}\int \Big(\mathbf{A}(\mathbf{r},t)
\mathbf{P}\Big)\mathbf{P}\times $$$$\times\Bigg[
\dfrac{2ie}{cp_T}\dfrac{\omega\tau-\mathbf{k}_1 \mathbf{P}}
{1-i \omega\tau +i\mathbf{k}_1 \mathbf{P}}g(P)+%$$$$+
\dfrac{iel}{c\hbar}\dfrac{f_F^+(\mathbf{P})-f_F^-(\mathbf{P})}
{1-i \omega\tau +i\mathbf{k}_1 \mathbf{P}}\Bigg]d^3P.
\eqno{(4.1)}
$$

We notice that
$$
\Big(\mathbf{P}\mp\dfrac{\hbar \mathbf{k}}{2p_T}\Big)=P^2\mp
\dfrac{\hbar}{p_T}\mathbf{Pk}+\Big(\dfrac{\hbar}{2p_T}\Big)^2k^2=
$$
$$
=P^2\mp\dfrac{\hbar \nu}{2\E_T}\mathbf{Pk}_1+\Big(\dfrac{\hbar \nu}
{4\E_T}\Big)^2k_1^2.
$$

Let's copy the previous equality in the form
$$
\mathbf{j}(\mathbf{r},t)=\dfrac{2ie^2p_T^4}{(2\pi\hbar)^3m}\int
[\mathbf{PA}(\mathbf{r},t)]\mathbf{P}S(P,\mathbf{Pk}_1)d^3P,
\eqno{(4.2)}
$$
where
$$
S(P,\mathbf{Pk}_1)= \dfrac{2}{cp_T}\dfrac{\omega\tau-\mathbf{k}_1 \mathbf{P}}
{1-i \omega\tau +i\mathbf{k}_1 \mathbf{P}}g(P)+ \dfrac{l}{c\hbar}\dfrac{f_F^+(\mathbf{P})-f_F^-(\mathbf{P})}
{1-i \omega\tau +i\mathbf{k}_1 \mathbf{P}}.
$$

We take the unit vector $\mathbf{e}_1=\dfrac{\mathbf{A}}{A}$, direct 
lengthwise the
vector $\mathbf{A}$. Then the equality (4.2) we may write in the form:
$$
\mathbf{j}(\mathbf{r},t)=\dfrac{2ie^2p_T^4A(\mathbf{r},t)}{(2\pi\hbar)^3m}\int
(\mathbf{Pe}_1)\mathbf{P}S(P,\mathbf{Pk}_1)d^3P,
\eqno{(4.3)}
$$

In view of the symmetry the value of
integral will not change, if the vector $\mathbf{e}_1$ is replaced
by any other unit vector $\mathbf{e}_2$, perpendicular to the vector
$\mathbf{k}_1$, i.e.
$$ \mathbf{e}_2=\dfrac{\mathbf{A} \times \mathbf{k}_1}
{|\mathbf{A} \times \mathbf{k}_1|}=
\dfrac{\mathbf{A} \times \mathbf{k}_1}{Ak_1},
$$
and $\mathbf{A}\times\mathbf{k}_1$ is the vector product.

Let's spread out a vector $\mathbf{P}$ in three orthogonal directions                                                           
$\mathbf{e}_1, \mathbf{e}_2$ and $\mathbf{n}=\dfrac{\mathbf{k}_1}{k_1}$:
$$
\mathbf{P}=(\mathbf{P n})\mathbf{n}+(\mathbf{Pe}_1)\mathbf{e}_1+
(\mathbf{Pe}_2)\mathbf{e}_2.
$$

By means of this decomposition we receive that
$$
(\mathbf{PA})\mathbf{P}=A(\mathbf{Pe}_1)\mathbf{P}=$$$$= A(\mathbf{Pe}_1)(\mathbf{Pn})\mathbf{n}+
A(\mathbf{Pe}_1)^2\mathbf{e}_1+A(\mathbf{Pe}_1)(\mathbf{Pe}_2)\mathbf{e}_2.
$$
Substituting this decomposition in (4.3), and, considering, 
that integrals from odd functions 
on a symmetric interval are equal to zero, we receive that
$$
\mathbf{j}(\mathbf{r},t)=\dfrac{2ie^2p_T^4\mathbf{A}(\mathbf{r},t)}
{(2\pi\hbar)^3m}\int (\mathbf{Pe}_1)^2S(P,\mathbf{Pn})d^3P,
\eqno{(4.4)}
$$
or, in the explicit form,
$$
\mathbf{j}(\mathbf{r},t)=
\dfrac{2ep_T^4\mathbf{A}(\mathbf{r},t)}{(2\pi \hbar)^3m}
\int \Big(\mathbf{e}_1
\mathbf{P}\Big)^2\Bigg[
\dfrac{2ie}{cp_T}\dfrac{\omega\tau-\mathbf{k}_1 \mathbf{P}}
{1-i \omega\tau +i\mathbf{k}_1 \mathbf{P}}g(P)+$$$$+
\dfrac{iel}{c\hbar}\dfrac{f_F^+(\mathbf{P})-f_F^-(\mathbf{P})}
{1-i \omega\tau +i\mathbf{k}_1 \mathbf{P}}\Bigg]d^3P.
$$

Here $\mathbf{e}_1=\mathbf{A}/A $ is the unit vector directed
lengthwise $\mathbf{A}$.  Therefore  in view of symmetry
$$
\int \Big(\mathbf{e}_1
\mathbf{P}\Big)^2[S]d^3P=
\int \Big(\mathbf{e}_2
\mathbf{P}\Big)^2[S]d^3P=
$$
$$
=\dfrac{1}{2}\int \Big[\Big(\mathbf{e}_1 \mathbf{P}\Big)^2
+\Big(\mathbf{e}_2 \mathbf{P}\Big)^2\Big]
[S]d^3P.
$$

We will notice that the square of length of a vector $\mathbf{P}$ is equal
$$
P^2=(\mathbf{Pe}_1)^2+(\mathbf{Pe}_2)^2+(\mathbf{Pn})^2,
$$
therefore
$$
(\mathbf{Pe}_1)^2+(\mathbf{Pe}_2)^2=P^2-\dfrac{(\mathbf{Pk}_1)^2}{k_1^2}=
P^2-(\mathbf{Pn})^2=P_\perp^2,
$$
where  $P_\perp$ is the projection of the vector $\mathbf{P}$ to direct, 
perpendicular planes $(\mathbf{e}_1,\mathbf{e}_2)$, and
the vector $\mathbf{n}$ is the unit vector directed along
the vector
$\mathbf{k}_1,\; \mathbf{n}=\dfrac{\mathbf{k}_1}{k_1}=\dfrac{\mathbf{k}}{k}$.

Hence for the current density we receive the following expression
$$
\mathbf{j}(\mathbf{r},t)=
\dfrac{ep_T^4\mathbf{A}(\mathbf{r},t)}{(2\pi \hbar)^3m}\int
\Big[P^2-(\mathbf{Pn})^2\Big]\Bigg[
\dfrac{2ie}{cp_T}\dfrac{\omega\tau-\mathbf{k}_1 \mathbf{P}}
{1-i \omega\tau +i\mathbf{k}_1 \mathbf{P}}g(P)+
$$
$$
+\dfrac{iel}{c\hbar}\dfrac{f_F^+(\mathbf{P})-f_F^-(\mathbf{P})}
{1-i \omega\tau +i\mathbf{k}_1 \mathbf{P}}\Bigg]d^3P.
$$

Replacing the current density in the left--hand side of this equality by the
expression in terms of field, we receive:
$$
\sigma_{tr}\dfrac{i \omega}{c}\mathbf{A}(\mathbf{r},t)=
%$$
%$$
%=
\dfrac{ep_T^4\mathbf{A}(\mathbf{r},t)}{(2\pi \hbar)^3m}\int \Big[
P^2-(\mathbf{Pn})^2\Big]\times
$$
$$
\times\Bigg[
\dfrac{2ie}{cp_T}\dfrac{\omega\tau-\mathbf{k}_1 \mathbf{P}}
{1-i \omega\tau +i\mathbf{k}_1 \mathbf{P}}g(P)+%$$$$+
\dfrac{iel}{c\hbar}\dfrac{f_F^+(\mathbf{P})-f_F^-(\mathbf{P})}
{1-i \omega\tau +i\mathbf{k}_1 \mathbf{P}}\Bigg]d^3P.
$$\medskip

\begin{center}
  \bf 5. ELECTRIC CONDUCTIVITY AND DIELECTRIC FUNCTION
\end{center}

From the last formula we receive the following expression for the
transversal electric conductivity in quantum non-degenerate plasma:
$$
\sigma_{tr}=
\dfrac{e^2p_T^4}{(2\pi \hbar)^3m \omega}\int \Big[
P^2-
\dfrac{(\mathbf{P}\mathbf{k}_1)^2}{k_1^2}\Big]\Bigg[
\dfrac{2}{p_T}\dfrac{\omega\tau-\mathbf{k}_1 \mathbf{P}}
{1-i \omega\tau +i\mathbf{k}_1 \mathbf{P}}g(P)+
$$
$$
+\dfrac{l}{\hbar}\dfrac{f_F^+(\mathbf{P})-f_F^-(\mathbf{P})}
{1-i \omega\tau +i\mathbf{k}_1 \mathbf{P}}\Bigg]d^3P.
\eqno{(5.1)}
$$

We will transform expression for transversal conductivity and
we will bring it to the form:
$$
\sigma_{tr}=
\dfrac{2e^2p_T^3}{(2\pi \hbar)^3 m \omega}\int \Bigg[
(\omega\tau-\mathbf{k}_1\mathbf{P})g(P)%+ \hspace{6cm}
%$$
%$$
%\hspace{1cm}
+\dfrac{\E_T}{\hbar \nu}(f_F^+-f_F^-)\Bigg]
\dfrac{P_\perp^2d^3P}{1-i \omega\tau +i\mathbf{k}_1
\mathbf{P}},
$$
where $\E_T$ is the thermal electron energy,
$$
\E_T=\dfrac{mv_T^2}{2}.
$$

With the use of the equality (2.3) we will present the previous
formula in the form:
$$
\sigma_{tr}=\dfrac{\sigma_0}{4\pi f_2(\alpha)}
\int\Bigg[\big[1-(\omega\tau)^{-1}\mathbf{k}_1\mathbf{P}\Big]g(P)+
\hspace{4cm}
$$
$$\hspace{2cm}
+\dfrac{\E_T}{\hbar \omega}
\Big[f_F^+(\mathbf{P})-f_F^-(\mathbf{P})\Big]\Bigg]
\dfrac{(P^2-(\mathbf{Pn})^2)\;d^3P}{1-i\omega\tau+i\mathbf{k}_1
\mathbf{P}}.
\eqno{(5.1')}
$$

Here the function $f_2(\alpha)$ has been entered above and in the
absence of the scalar potential it is defined by equality:
$$
f_2(\alpha)=\int\limits_{0}^{\infty}x^2f_F(x)dx=\int\limits_{0}^{\infty}
\dfrac{x^2\,dx}{1+e^{x^2-\alpha}}=\dfrac{1}{2}\int\limits_{0}^{\infty}
\ln(1+e^{\alpha-x^2})dx.
$$

The quantity $\sigma_0$ is defined by classical expression for the
static electric conductivity
$$
\sigma_0=\dfrac{e^2N^{(0)}}{m\nu}.
$$

Dielectric function we will find according to the formula:
$$
\varepsilon_{tr}=
1+\frac{4\pi i}{\omega}\sigma_{tr}.
$$

Substituting electric conductivity (3.1) into this equality, we
receive the expression for dielectric permittivity in quantum  non-degenerate
collision plasma:
$$
\varepsilon_{tr}=1+\dfrac{\omega_p^2}{\omega^2}
 \dfrac{i}{4\pi f_2(\alpha)}
\int\Bigg[\Big[\omega\tau-\mathbf{k}_1\mathbf{P}\Big]g(P)+
\hspace{4cm}
$$
$$\hspace{2cm}
+\dfrac{\E_T}{\hbar \nu}\Big[f_F^+(\mathbf{P})-f_F^-(\mathbf{P})\Big]\Bigg]
\dfrac{P_\perp^2\;d^3P}{1-i\omega \tau+i\mathbf{k}_1\mathbf{P}}.
\eqno{(5.2)}
$$

We investigate some special cases of electroconductivity. In the
long-wave limit (when $k\to 0$) from (3.1) we receive the well known
classical expression:
$$
\sigma_{tr}(k=0)=\sigma_0\dfrac{\nu}{\nu-i \omega}=
\dfrac{\sigma_0}{1-i\omega\tau}.
$$

Let's consider the quantum mechanical limit of the conductivity in
the case of arbitrary values of wave number, i.e. conductivity limit
in the case, when Planck's constant $\hbar\to 0$, and the quantity
$k$ is arbitrary.

Now we consider the case, when values of the wave number are
arbitrary, but Planck's constant converges to zero: $\hbar\to 0$.

When the values of $\hbar$ are small we have:
$$
f_0^{\pm}(\mathbf{P})=f_F(P)\pm g(P)2\mathbf{P}\dfrac{\hbar
\mathbf{k}}{2mv_T},
$$
hence
$$
f_F^+(\mathbf{P})-f_F^-(\mathbf{P})=2g(P)2\mathbf{P}\dfrac{\hbar
\mathbf{k}}{2mv_T}.
$$

Therefore
$$
(\omega-v_T\mathbf{k}\mathbf{P})g(P)
+\dfrac{p_T^2}{2m\hbar}\big[f_F^+(\mathbf{P})-f_F^-(\mathbf{P})\big]=
\omega g(P).
$$

Thus, in linear approximation at small $\hbar$ (independently of the
quantity $k$) for transverse conductivity we receive:
$$
\sigma_{tr}=\sigma_{tr}^{\rm classic},
$$
where
$$
\sigma_{tr}^{\rm classic}=\dfrac{\sigma_0}{4\pi f_2(\alpha)}\int
\dfrac{g(P)P_\perp^2\,d^3P}{1-i\omega \tau +i
\mathbf{k}_1\mathbf{P}}.
\eqno{(5.3)}
$$

The expression (5.3) accurately coincides with the expression of the
trans\-ver\-sal conductivity for classical plasma with arbitrary
temperature.

Let's return to the expression $(5.1')$. We present it in the form of
the sum of two components
$$
\sigma_{tr}=\sigma_{tr}^{\rm classic}+\sigma_{tr}^{\rm quant},
\eqno{(5.4)}
$$
where $\sigma^{\rm classic}$ is defined by the equality (5.3), and
second component $\sigma_{tr}^{\rm quant}$ corresponds to quantum
properties of the plasma under consideration
$$
\sigma_{tr}^{\rm quant}=\dfrac{\sigma_0}{4\pi f_2(\alpha)}
\int\Bigg[-\dfrac{\mathbf{k}_1\mathbf{P}}{\omega\tau}g(P)+\hspace{5cm}$$$$
\hspace{3cm}+
\dfrac{\E_T}{\hbar \omega}[f_F^+(\mathbf{P})-f_F^-(\mathbf{P})]\Bigg]
\dfrac{P_\perp^2\,d^3P}{1-i\omega\tau+i\mathbf{k}_1\mathbf{P}}.
\eqno{(5.5)}
$$

The quantum summand $\sigma_{tr}^{\rm quant}$ we will present in the
form, proportional to a square of the Planck's constant $\hbar$.

For this aim we use cubic expansion of $\sigma_{tr}^{\rm quant}$
by powers of $\hbar$. We will remind, that in linear approximation
by $\hbar$, as it was already specified, the quantity
$\sigma_{tr}^{\rm quant}$ disappears.
We will direct an axis $x$ along the wave vector $\mathbf{k}$.

Let's expand the Fermi --- Dirac distribution by degrees of
dimensionless wave number $q=\dfrac{k}{k_T}=\dfrac{k_1\hbar\nu}
{mv_T^2}$, where $k_T=\dfrac{p_T}{\hbar}$ is the thermal wave
number. We receive that
$$
f_F^{\pm}(\mathbf{P})=f_F(P)\pm g(P)P_x q-
\Big[g'_{P^2}(P)P_x^2+\dfrac{1}{2}
g(P)\Big]\dfrac{q^2}{2}\pm
$$
$$
\pm\Big[g''_{P^2P^2}(P)P_x^2+
\dfrac{3}{2}g'_{P^2}(P)\Big]P_x\dfrac{q^3}{6}+\cdots.
$$

Here \quad$g'_{P^2}(P)=g'(P^2),\;g''_{P^2P^2}(P)=g''(P^2)$,
$$
g'(P^2)=g(P)\dfrac{1-e^{P^2-\alpha}}{1+e^{P^2-\alpha}},\quad
g''(P^2)=g(P)\Big[\Big(\dfrac{1-e^{P^2-\alpha}}{1+e^{P^2-\alpha}}\Big)^2-
2g(P)\Big].
$$

Now it is easy to find the difference
$$
f_F^+(\mathbf{P})-f_F^-(\mathbf{P})=
2g(P)P_xq+\Big[g''(P^2)P_x^2+
\dfrac{3}{2}g'(P^2)\Big]P_x\dfrac{q^3}{3}+\cdots.
$$

By means of this expression we find, that
$$
-\dfrac{k_1P_x}{\omega\tau}g(P)+\dfrac{\E_T}{\hbar\omega}
[f_F^+(\mathbf{P})-f_F^-(\mathbf{P})]=
G(\mathbf{P})\dfrac{k_1^3\hbar^2 \nu^3}{6\omega
m^2v_T^4}+\cdots,
$$
where
$$
G(\mathbf{P})=P_x\Big[g''(P^2)P_x^2+
\dfrac{3}{2}g'(P^2)\Big].
$$

Substituting this expression into (5.5), we obtain, that the quantum
summand is proportional to the square of Planck's constant and it is
defined by expression
$$
\sigma_{tr}^{\rm quant}=\hbar^2 \sigma_0
\dfrac{k_1^3 \nu^3}{24\pi \omega m^2v_T^4 f_2(\alpha)}
\int\dfrac{G(\mathbf{P})(P^2-P_x^2)\,d^3P}{1-i\omega\tau+ik_1P_x}.
$$

Let's similarly (5.4) we present 
the formula for calculation of dielectric permeability in the form
$$
\varepsilon_{tr}=1+\dfrac{4\pi i}{\omega}\Big(\sigma_{tr}^{\rm classic}+
\sigma_{tr}^{\rm quant}\Big)=\varepsilon_{tr}^{\rm classic}+
\varepsilon_{tr}^{\rm quant},
\eqno{(5.6)}
$$
where
$$
\varepsilon_{tr}^{\rm classic}=1+
\dfrac{4\pi i}{\omega}\sigma_{tr}^{\rm classic}, \qquad
\varepsilon_{tr}^{\rm quant}=\dfrac{4\pi i}{\omega}
\sigma_{tr}^{\rm quant}.
$$

Thus, in an explicit form dielectric 
permeability of classical non-degenerate collisional plasmas is equal  
$$
\varepsilon_{tr}^{\rm classic}=1+i\dfrac{\omega_p^2}{\omega^2}\cdot
\dfrac{\omega\tau}{4\pi f_2(\alpha)}\int \dfrac{g(P)P_\perp^2d^3P}
{1-i\omega\tau+i\mathbf{k}_1\mathbf{P}}. \eqno{(5.7)}
$$
And the component of dielectric permeability answering to quantum properties
of  non-degenerate collisional plasmas is defined by equality $$
\varepsilon_{tr}^{\rm quant}=i\dfrac{\omega_p^2}{\omega^2}\cdot
\dfrac{1}{4\pi f_2(\alpha)}\int\dfrac{-\mathbf{k_1P}g(P)+(\E_T/\hbar \nu)
(f_F^+-f_F^-)}{1-i\omega\tau+i\mathbf{k}_1\mathbf{P}}.
\eqno{(5.8)}
$$

\begin{center}
\bf 6. CALCULATION OF ELECTRIC CONDUCTIVITY AND DIELECTRIC
PERMEABILITY

\end{center}

In the expressions for classical and quantum components of the
conductivity we can simplify several integrals.

We break the triple integral to external one--dimensional integration
by the variable $P_x$ from $-\infty$ to $+\infty$ and internal
double integration by plane orthogonal to the axis $P_x$ in the
expression (5.3). The internal integration we carry out in polar
coordinates. Here we obtain that
$$
P^2=P_x+P_{\perp}^2, \qquad d^3P=dP_x\;d{\mathbf{P_{\perp}}}, \qquad
d{\mathbf{P_{\perp}}}=P_{\perp}\,dP_{\perp}d\chi,
$$
where $P_{\perp}$ is the polar radius, and $\chi$ is the polar angle.

Thus we receive, that
$$
\sigma_{tr}^{\rm classic}=\dfrac{\sigma_0}{4\pi f_2(\alpha)}
\int\limits_{-\infty}^{\infty}dP_x
\int\limits_{0}^{\infty}\int\limits_{0}^{2\pi}
\dfrac{g(P)\,P_{\perp}^3 dP_{\perp} d\chi}{1-i\omega \tau
+ik_1P_x},
$$
where
$$
g(P)=\dfrac{e^{P_x^2+P_{\perp}-\alpha}}{(1+e^{P_x^2+P_{\perp}-\alpha})^2}.
$$

Internal double integral we calculate in polar coordinates
$$
\int\limits_{0}^{\infty}\int\limits_{0}^{2\pi}
g(P)\,P_{\perp}^3 dP_{\perp} d\chi=\pi\ln(1+e^{\alpha-P_x^2}).
\eqno{(6.1)}
$$

Hence, the expression for the classical component is simplified to
one-dimensional integral
$$
\sigma_{tr}^{\rm classic}=
\dfrac{\sigma_0}{4f_2(\alpha)}
\int\limits_{-\infty}^{\infty}\dfrac{\ln(1+e^{\alpha-
P_x^2})dP_x}{1-i\omega\tau+ik_1P_x}.
\eqno{(6.2)}
$$

The quantum item (5.5) we present in the form of the sum of two
items
$$
\sigma_{tr}^{\rm quant}=\sigma_1+\sigma_2.
\eqno{(6.3)}
$$

Here
$$
\sigma_1=-\dfrac{\sigma_0 k_1}{4\pi f_2(\alpha)\omega\tau}\int
\dfrac{P_x(P^2-P_x^2)g(P)\,d^3P}{1-i\omega\tau +ik_1P_x},
$$
and
$$
\sigma_2=\dfrac{\sigma_0\E_T}
{4\pi f_2(\alpha)\hbar \omega}\int
\dfrac{f_F^+(\mathbf{P})-f_F^-(\mathbf{P})}{1-i\omega\tau+ik_1P_x}
(P^2-P_x^2)d^3P.
\eqno{(6.4)}
$$

With the help of the equality (6.1) the expression for $\sigma_1$
can be rewritten in the following form
$$
\sigma_1=-\dfrac{\sigma_0 k_1}{4f_2(\alpha)\omega\tau}\int
\limits_{-\infty}^{\infty}\dfrac{P_x\ln(1+e^{\alpha-
P_x^2})dP_x}{1-i\omega\tau+ik_1P_x}.
\eqno{(6.5)}
$$

After change of variable
$$
P_x\mp \dfrac{\hbar k}{2p_T}\equiv P_x\mp \dfrac{k_1\hbar \nu}
{2mv_T^2}\equiv P_x\mp \dfrac{k_1\hbar \nu}{4\E_T} \to P_x
$$
the difference of integrals from (6.4) will be transformed
to one integral and we receive \smallskip
$$
\sigma_2=-\dfrac{i\sigma_0k_1^2}{8\pi f_2(\alpha)\omega\tau}
\int
\dfrac{f_F(P)(P^2-P_x^2)\,d^3P}{(1-i\omega\tau+ik_1P_x)^2+
(k_1^2\hbar\nu/4\E_T)^2}.
\eqno{(6.6)}
$$
\medskip

In the same way, as well as during the derivation of the formula
(6.1), double internal integral in (6.6) we reduce to the
one-dimensional integral
$$
\int\limits_{-\infty}^{\infty}\int\limits_{-\infty}^{\infty}
f_F(P)[P^2-P_x^2]\;d\,\mathbf{P}_{\perp}=
\int\limits_{0}^{\infty}\int\limits_{0}^{2\pi}
f_F(P)P_{\perp}^3\,dP_{\perp}d\chi=
$$
$$
=2\pi \int\limits_{0}^{\infty}
{P}_{\perp}\ln(1+e^{\alpha-P_x^2-P_{\perp}^2})\,d{P}_{\perp}.
$$

Now the expression (6.6) can be written in the following form
(replacing a variable of integration $P_{\perp}=\rho$)
$$
\sigma_2=-\dfrac{i\sigma_0 k_1^2}{4f_2(\alpha)\omega\tau}
\int\limits_{-\infty}^{\infty}\int\limits_{0}^{\infty}
\dfrac{\rho\ln(1+e^{\alpha-\rho^2-P_x^2})d\rho\,dP_x}
{(1-i\omega\tau+ik_1P_x)^2+(k_1^2\hbar\nu/4\E_T)^2}.
\eqno{(6.7)}
$$

Thus, it is possible to present the expression for transverse
conductivity in the form of the sum of one-dimensional (6.2), (6.5) and
two-dimensional (6.7) integrals
$$
\sigma_{tr}=\dfrac{\sigma_0}{4f_2(\alpha)}
\int\limits_{-\infty}^{\infty}\dfrac{[1-(k_1/\omega\tau)P_x]\ln(1+e^{\alpha-
P_x^2})dP_x}{1-i\omega\tau+ik_1P_x}-
$$
$$
-\dfrac{i\sigma_0 k_1^2}{4 f_2(\alpha)\omega \tau}
\int\limits_{-\infty}^{\infty}\int\limits_{0}^{\infty}
\dfrac{\rho\ln(1+e^{\alpha-\rho^2-P_x^2})d\rho\,dP_x}
{(1-i\omega\tau+ik_1P_x)^2+(k_1^2\hbar\nu/4\E_T)^2}.
$$

In the expression (6.6) for $\sigma_2$ the thriple integral can be
reduced to one-dimensional integral. For this purpose in (6.5) we
pass to integration in spherical coordinates and present this
expression in the form
$$
\sigma_2=-
\dfrac{i\sigma_0k_1^2}{4f_2(\alpha)\omega\tau}
\int\limits_{0}^{\infty} f_F(P)P^4 J(P)\,dP,
$$
where
$$
J(P)=\int\limits_{-1}^{1}\dfrac{(1-\mu^2)\;d\mu}
{(1-i\omega\tau +ik_1P\mu)^2+(k_1^2\hbar\nu/4\E_T)^2}.
$$

Let's designate temporarily
$$
a=1-i\omega\tau, \qquad b=ik_1P, \qquad d=\dfrac{\hbar \nu k_1^2}{4\E_T},
$$
and rewrite the integral $J(P)$ in the form:
$$
J= \int\limits_{-1}^{1}\dfrac{(1-\mu^2)d\mu}{(a+b\mu)^2+d^2}.
$$

After change of variable $a+b\mu=t $ this integral will be rewritten
in the form
$$
J=\dfrac{1}{b^3}\int\limits_{a-b}^{a+b}\dfrac{b^2-(t-a)^2}{t^2+d^2}dt.
$$

This integral equals to
$$
J=-\dfrac{2}{b^2}+\dfrac{d^2+b^2-a^2}{b^3} \dfrac{1}{2id}
\ln\dfrac{(a+b-id)(a-b+id)}{(a+b+id)(a-b-id)}+
$$
$$
+\dfrac{a}{b^3}\ln\dfrac{(a+b-id)(a+b+id)}{(a-b-id)(a-b+id)},
$$
or
$$
J=-\dfrac{2}{b^2}+\dfrac{d^2+b^2-a^2}{2idb^3}\ln\dfrac{a^2-(b-id)^2}
{a^2-(d+id)^2}+\dfrac{a}{b^3}\ln\dfrac{(a+b)^2+d^2}
{(a-d)^2+d^2}.
$$

Considering designations for $a,b,d$, we receive
$$
J(P)\equiv J(P;\omega\tau,k_1)=\dfrac{2}{(k_1P)^2}-
$$$$-\dfrac{(1-i\omega\tau)^2+(k_1P)^2-
(\hbar \nu k_1^2/4\E_T)^2}{k_1^5P^3(\hbar \nu/2\E_T)}
\ln\dfrac{(1-i\omega\tau)^2+(k_1P-\hbar \nu k_1^2/4\E_T)^2}
{(1-i\omega\tau)^2+(k_1P+\hbar \nu k_1^2/4\E_T)^2}+
$$
$$
+i\dfrac{1-i\omega\tau}{(k_1P)^3}\ln
\dfrac{(1-i\omega\tau+ik_1P)^2+(\hbar \nu k_1^2/4\E_T)^2}
{(1-i\omega\tau-ik_1P)^2+(\hbar \nu k_1^2/4\E_T)^2}.
\eqno{(6.8)}
$$

Thus, the expression of quantum transverse conductivity is defined
by one-dimensional integral
$$
\sigma_{tr}=\dfrac{\sigma_0}{4f_2(\alpha)}
\int\limits_{-\infty}^{\infty}\dfrac{1-(k_1/\omega\tau)P_x}
{1-i\omega\tau+ik_1P_x}\ln(1+e^{\alpha-P_x^2})dP_x-$$$$-
\dfrac{i\sigma_0k_1^2}{4f_2(\alpha)\omega\tau}
\int\limits_{0}^{\infty} f_F(P)P^4 J(P)\,dP,%\qquad k_1=kl,
$$
where the function $J(P)$ is defined by expression (6.8).

So, for electric conductivity and 
dielectric function we have received following expressions
$$
\sigma_{tr}=\dfrac{\sigma_0}{4if_2(\alpha)}\Bigg[\int\limits_{-\infty}^{\infty}
\Big(\dfrac{1}{q}-\dfrac{\tau}{x}\Big)\dfrac{y\ln(1+e^{\alpha-\tau^2})d\tau}
{\tau-z/q}-\dfrac{y}{x}\int\limits_{0}^{\infty}
\dfrac{u^4J(u)du}{1+e^{u^2-\alpha}}\Bigg]
\eqno{(6.9)}
$$ and
$$
\varepsilon_{tr}=1+\dfrac{x_p^2}{4f_2(\alpha)x^2}\Bigg[
\int\limits_{-\infty}^{\infty}\dfrac{(x-q\tau)\ln(1+e^{\alpha-\tau^2})}
{\tau-z/q}d\tau-q\int\limits_{0}^{\infty}
\dfrac{u^4J(u)du}{1+e^{u^2-\alpha}}\Bigg].
\eqno{(6.10)}
$$

In formulas (6.9) and (6.10) following designations are accepted
$$
q=\dfrac{k}{k_T},\quad z=x+iy=\dfrac{\omega+i \nu}{k_Tv_T},\quad
x=\dfrac{\omega}{k_Tv_T},\quad y=\dfrac{\nu}{k_Tv_T},\quad x_p=\dfrac{\omega_p}
{k_Tv_T},
$$
$$
J(u)=-\dfrac{2}{u^2}-\dfrac{(z/q)^2+(q/2)^2-u^2}{u^3q}
\ln\dfrac{(u-q/2)^2-(z/q)^2}{(u+z/q)^2-(q/2)^2}-
$$
$$
-\dfrac{z}{u^3q}\ln\dfrac{(u-z/q)^2-(q/2)^2}{(u+z/q)^2-(q/2)^2}=
\int\limits_{-1}^{1}\dfrac{(1-\tau^2)d\tau}{(u\tau-z/q)^2-(q/2)^2}.
$$

Besides these formulas (6.9) and (6.10) we can give and other formulas for
electric conductivity and dielectric function
$$
\sigma_{tr}=\dfrac{\sigma_0}{4if_2(\alpha)}\int\limits_{-\infty}^{\infty}
\Big(\dfrac{y}{q}-\dfrac{y\tau}{x}\Big)\dfrac{\ln(1+e^{\alpha-\tau^2})}
{\tau-z/q}d\tau-
$$ 
$$
-\dfrac{y\sigma_0}{4if_2(\alpha)x}\int\limits_{-\infty}^{\infty}
\dfrac{d\tau}{(\tau-z/q)^2-(q/2)^2}\int\limits_{0}^{\infty}
\rho\ln(1+e^{\alpha-\tau^2-\rho^2})d\rho
$$
and
$$
\varepsilon_{tr}=1+\dfrac{x_p^2}{4f_2(\alpha)x^2}\dfrac{x}{q}
\int\limits_{-\infty}^{\infty}
\Big(1-\dfrac{q}{x}\tau\Big)\dfrac{\ln(1+e^{\alpha-\tau^2})d\tau}{\tau-z/q}-
$$$$
-\dfrac{x_p^2}{4f_2(\alpha)x^2}\int\limits_{-\infty}^{\infty}
\dfrac{d\tau}{(\tau-z/q)^2-(q/2)^2}\int\limits_{0}^{\infty}
\rho\ln(1+e^{\alpha-\rho^2-\tau^2})d\rho.
$$
 
\begin{center}
\bf 7. THE SUM RULES
\end{center}
 
Let's check up performance of one of the parities, named the rule $f$-sums
(see, for example, \cite {Pains}) for transversal dielectric permeability
(6.14). This rule is expressed by the formula (4.200) of the monography 
\cite {Pains}
$$
\int\limits_{-\infty}^{\infty}\varepsilon_{tr}(q,\omega,\nu)\omega d\omega=
\pi \omega_p^2.
\eqno{(7.1)}
$$

As shown in \cite{Pains}, for the proof of the parity (7.1) it is enough
to prove performance of the limiting parity
$$
\varepsilon_{tr}(q,\omega,\nu)=1-\dfrac{\omega_p^2}{\omega^2}+
o\Big(\dfrac{1}{\omega^2}\Big),\qquad \omega\to \infty.
\eqno{(7.2)}
$$

From expression (6.10) it is visible that
$$
\varepsilon_{tr}^{\rm quant}=o\Big(\dfrac{1}{\omega^2}\Big),\qquad
\omega\to\infty.
$$
By means of this limiting parity we will present the parity (6.10) at big $ \omega $ ($ \omega \gg 1$) in the form
$$
\varepsilon_{tr}=1+\dfrac{i\omega_p^2 \omega \tau}{4\pi f_2(\alpha)\omega^2}
\int \dfrac{g(P)P_\perp^2\;d^3P}{1-i\omega \tau+i\mathbf{k}_1\mathbf{P}}.
\eqno{(7.3)}
$$

From comparison (7.2) and (7.3) it is visible, that now it is required 
to prove equality
$$
\dfrac{i\omega\tau}{4\pi f_2(\alpha)}
\int \dfrac{g(P)P_\perp^2\;d^3P}{1-i\omega \tau+i\mathbf{k}_1\mathbf{P}}=-1,
$$
or, considering once again, that $ \omega \gg 1$, is required to prove equality
$$
\dfrac{1}{4\pi f_2(\alpha)}
\int g(P)P_\perp^2\;d^3P=1.
\eqno{(7.4)}
$$

Passing in equality (7.4) to spherical coordinates and integrating
once in parts, we receive
$$
\dfrac{2}{3f_2(\alpha)}\int\limits_{0}^{\infty}\dfrac{e^{P^2-\alpha}P^4dP}
{(1+e^{P^2-\alpha})^2}=1.
\eqno{(7.5)}
$$

Of justice of equality (7.5) we are convinced, having integrated once by
parts.

\begin{center}
\bf 8. THE ANALYSIS OF RESULTS AND THE CONCLUSION
\end{center}

The analysis of graphics on fig. 1--7 shows, that at great values $q $ the 
quantum conductivity does not decrease to zero (as classical conductivity), 
and tends for collisional plasma to the finite limit. Really, from the 
formula (6.13) follows, that
$$
 \lim\limits_{q\to \infty}\sigma_{tr}=i\dfrac{y}{x}=i\dfrac{\nu}{\omega}.
\eqno{(8.1)}
$$

On Figs 1--7 all curves 1 answer to classical plasma, and curves 2 
to the quantum plasma.

On Figs. 1--3 are presented dependence $ |\sigma_{tr}/\sigma_0| $
quantum and classical conductivity from quantity $q=k/k_T $ at the various
values of the resulted chemical potential $ \alpha $: $ \alpha=0$ (fig. 1),
$ \alpha=6$ (fig. 2) and $ \alpha =-5$ (fig. 3). 

The analysis of graphics and numerical
calculations show at negative values of chemical potential
weak dependence $ |\sigma_{tr}/\sigma_0| $ on quantity  $q $, and strong
dependence at positive values of chemical potential.
From fig. 1--3 it is visible, that while the module of classical conductivity
has only one maximum, the module of quantum conductivity has also a minimum.
The range of values the module quantum conductivity decreases at decrease
chemical potential.

On fig. 4 and 5 dependence $ \Re(\sigma_{tr}/\sigma_0) $ is presented for
quantum and classical conductivity. From the parity (8.1) it is visible, that
the real part of quantum conductivity tends to zero at $q\to \infty $
the same as also the real part of classical conductivity. However,
decrease of quantum conductivity occurs much faster, than
decrease of classical conductivity that is visible from fig. 4 and 5. 
At the real parts of quantum conductivity there is a maximum, 
the minimum is absent.

On fig. 6 and 7 dependence $ \Im(\sigma_{tr}/\sigma_0) $ is presented
quantum and classical conductivity on quantity  $q $. Qualitatively 
imaginary part behaves the same as also the module. 
From fig. 1 - 7 it is visible, that at $q\to \infty $
quantum and classical conductivity coincide.

On fig. 8 dependence graphics $ |\sigma_{tr}/\sigma_0| $ are presented
quantum plasma from
quantities $q $ at various values of quantity $x $  (dimensionless frequency
oscillations of vector potential). The analysis of graphics shows, that 
with growth
quantity $x $ the range of values of the module of quantum plasma decreases, 
thus the maximum and module minimum are levelled, i.e. values of the module 
in points maximum and minimum approach, reducing range of values of the module 
of conductivity.

In the present work the correct formula for calculation of
transversal electric conductivity in the quantum non-degenerate collisinal
plasma is deduced.
For this purpose the Wigner---Vlasov---Boltzmann kinetic
equation with collisional integral in the form of BGK--model
(Bhatnagar, Gross and Krook) in coordinate space is used.

\begin{figure}[t]
\begin{center}
\includegraphics[width=16cm, height=10cm]{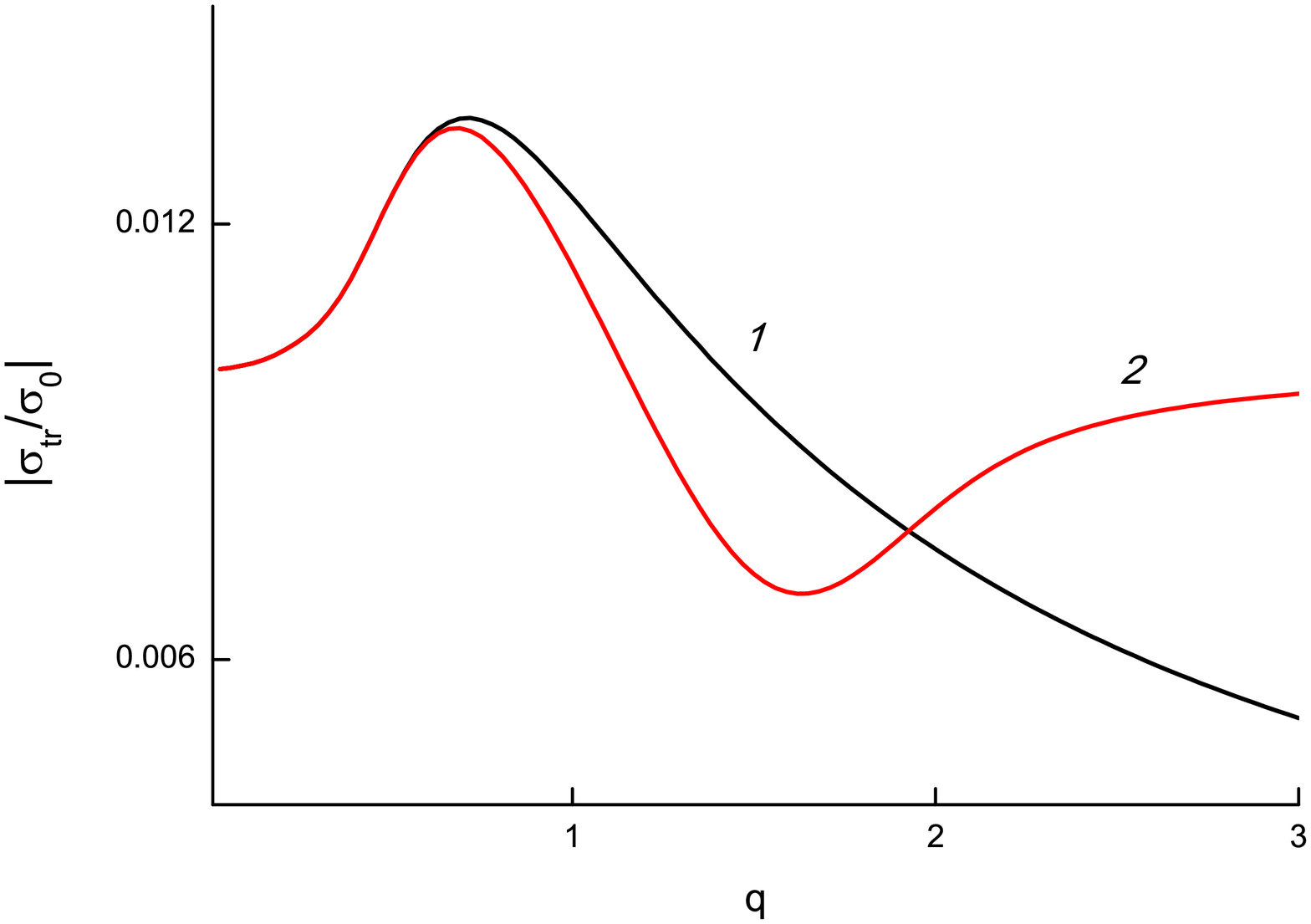}
\caption{Case: $x=1, y=0.01, \alpha=0.$
Dependence $|\sigma_{tr}/\sigma_0|$ on quantity $q$.}
\end{center}
%\end{figure}
%\begin{figure}[h]
\begin{center}
\includegraphics[width=16cm, height=10cm]{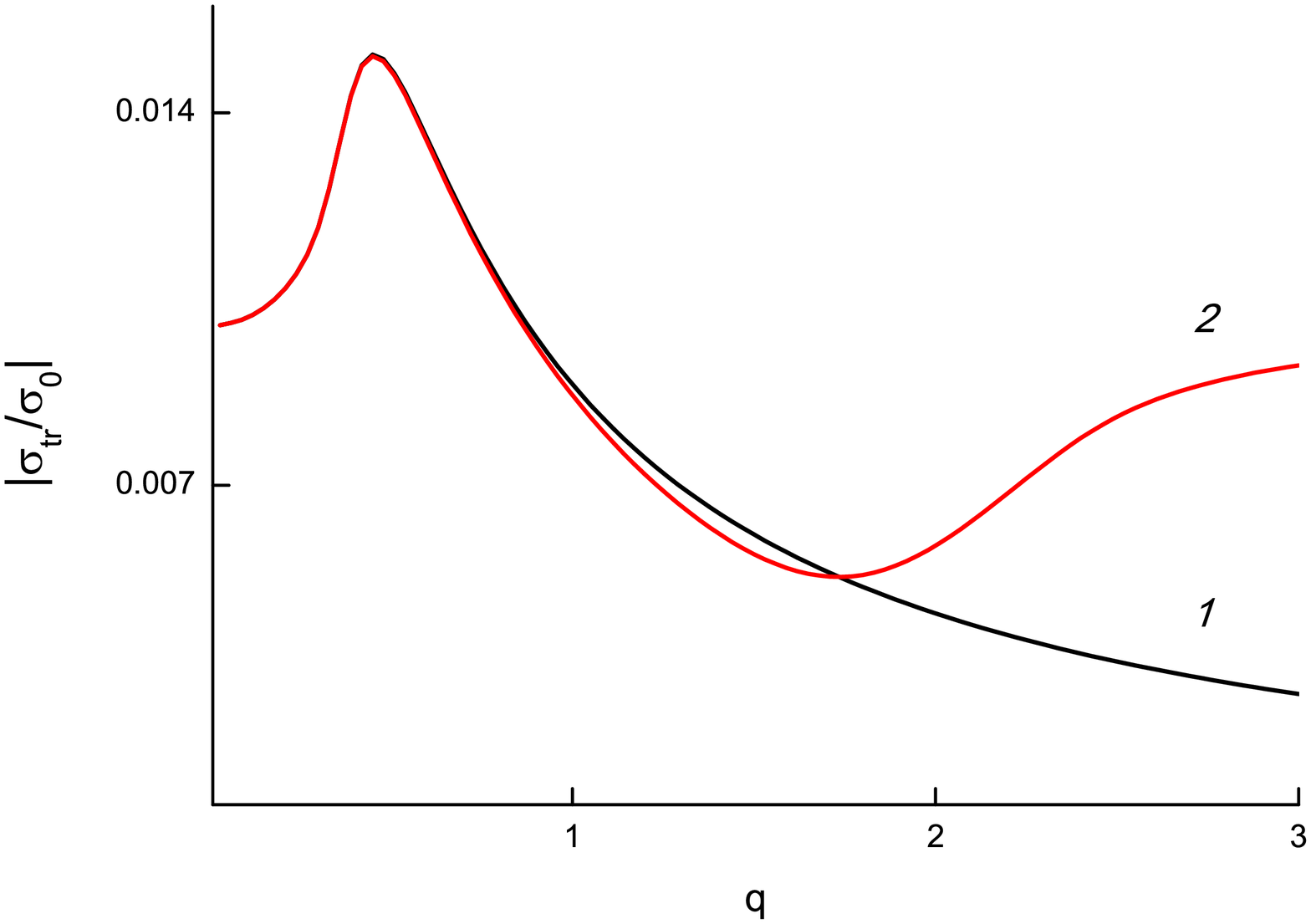}
\caption{Case: $x=1, y=0.01, \alpha=6.$
Dependence $|\sigma_{tr}/\sigma_0|$ on quantity $q$.}
\end{center}
\end{figure}

\begin{figure}[t]
\begin{center}
\includegraphics[width=16cm, height=10cm]{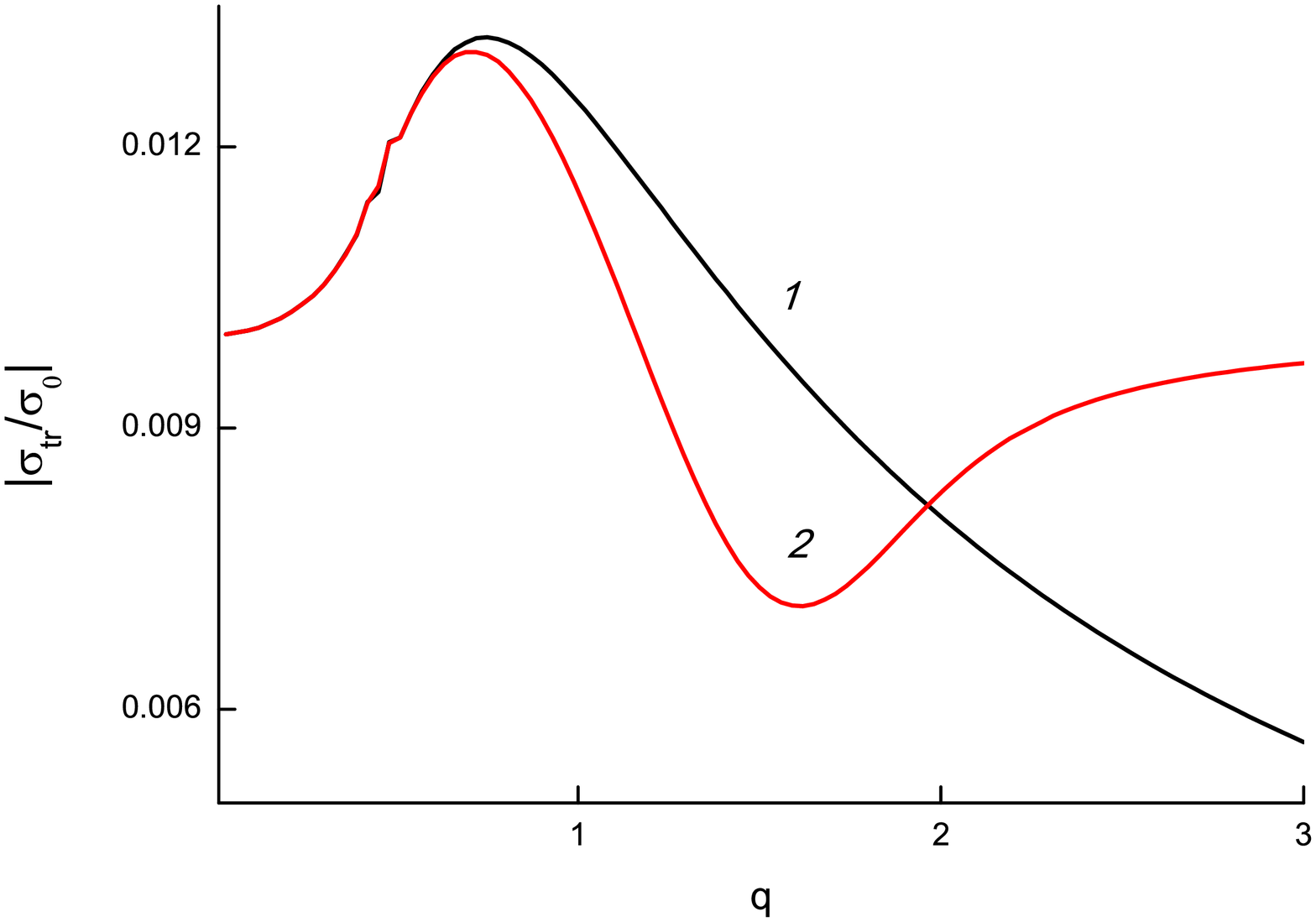}
\caption{Case: $x=1, y=0.01, \alpha=-5.$
Dependence $|\sigma_{tr}/\sigma_0|$ on quantity $q$.}
\end{center}
%\end{figure}
%\begin{figure}[h]
\begin{center}
\includegraphics[width=16cm, height=10cm]{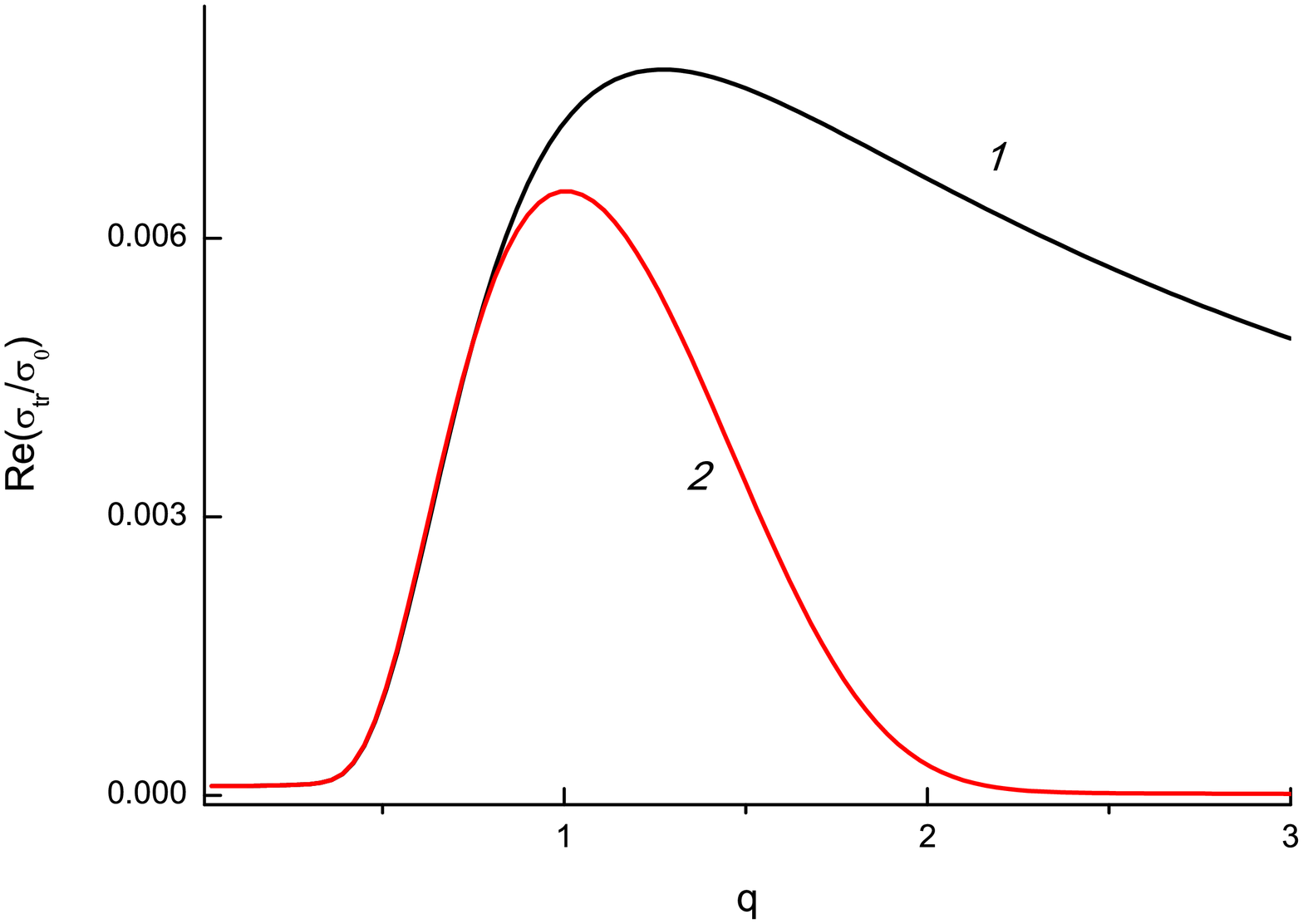}
\caption{Case: $x=1, y=0.01, \alpha=0.$
Dependence $\Re(\sigma_{tr}/\sigma_0)$ on quantity $q$.}
\end{center}
\end{figure}

\begin{figure}[t]
\begin{center}
\includegraphics[width=16cm, height=10cm]{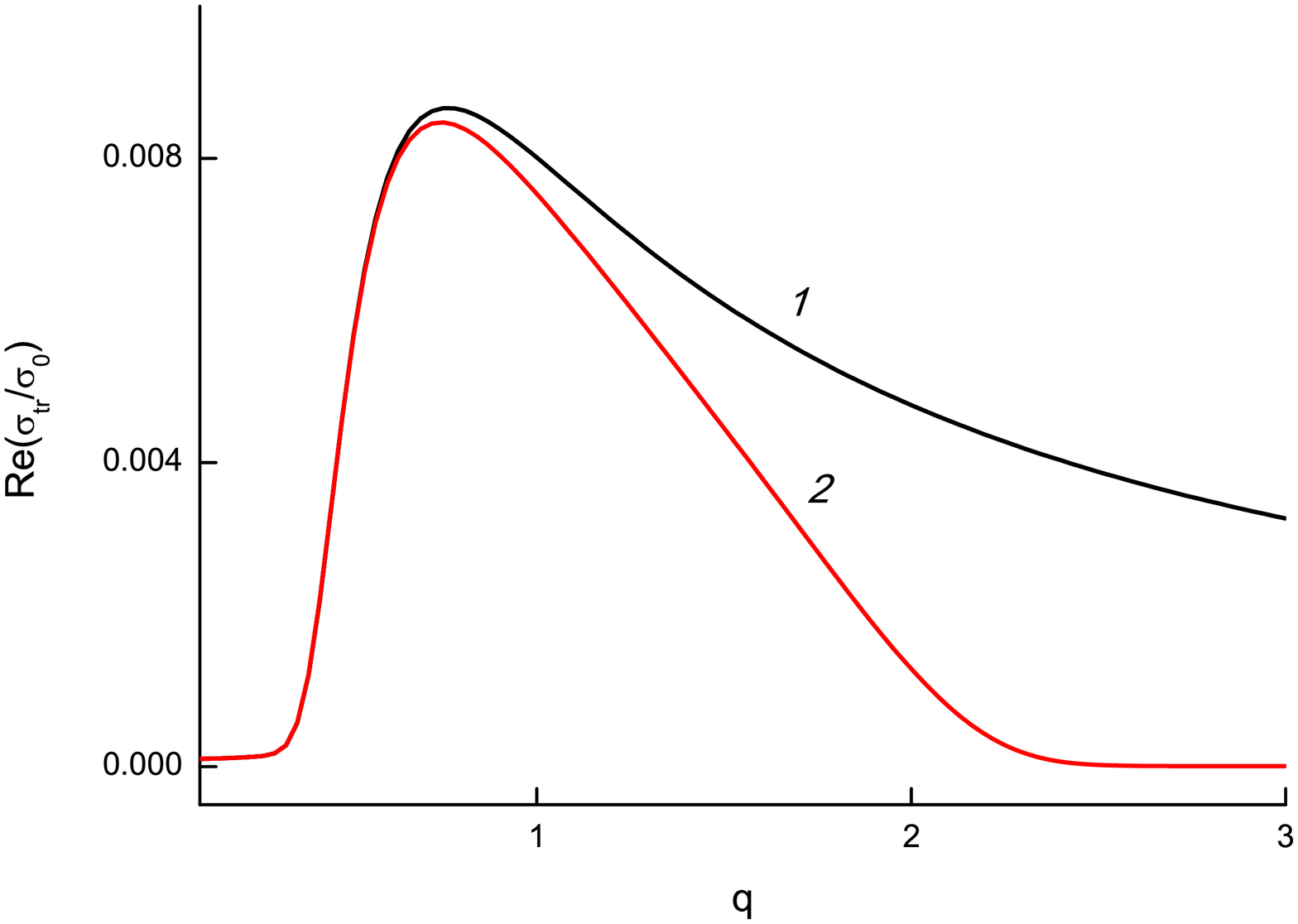}
\caption{Case: $x=1, y=0.01, \alpha=5.$
Pependence $\Re(\sigma_{tr}/\sigma_0)$ on quantity $q$.}
\end{center}
%\end{figure}
%\begin{figure}[h]
\begin{center}
\includegraphics[width=16cm, height=10cm]{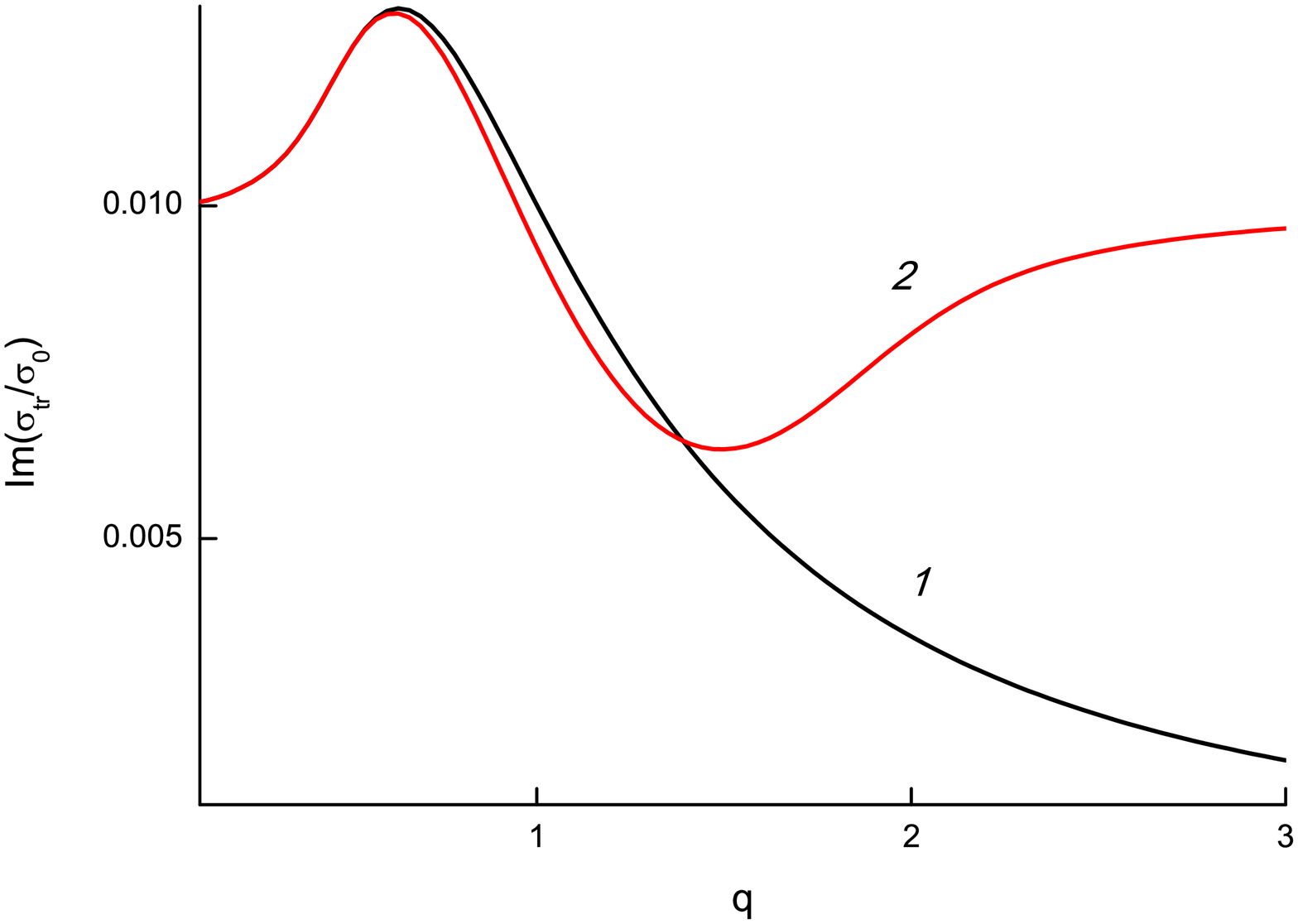}
\caption{Case: $x=1, y=0.01, \alpha=0.$
Dependence $\Im(\sigma_{tr}/\sigma_0)$ on quantity $q$.}
\end{center}
\end{figure}

\begin{figure}[t]
\begin{center}
\includegraphics[width=16cm, height=10cm]{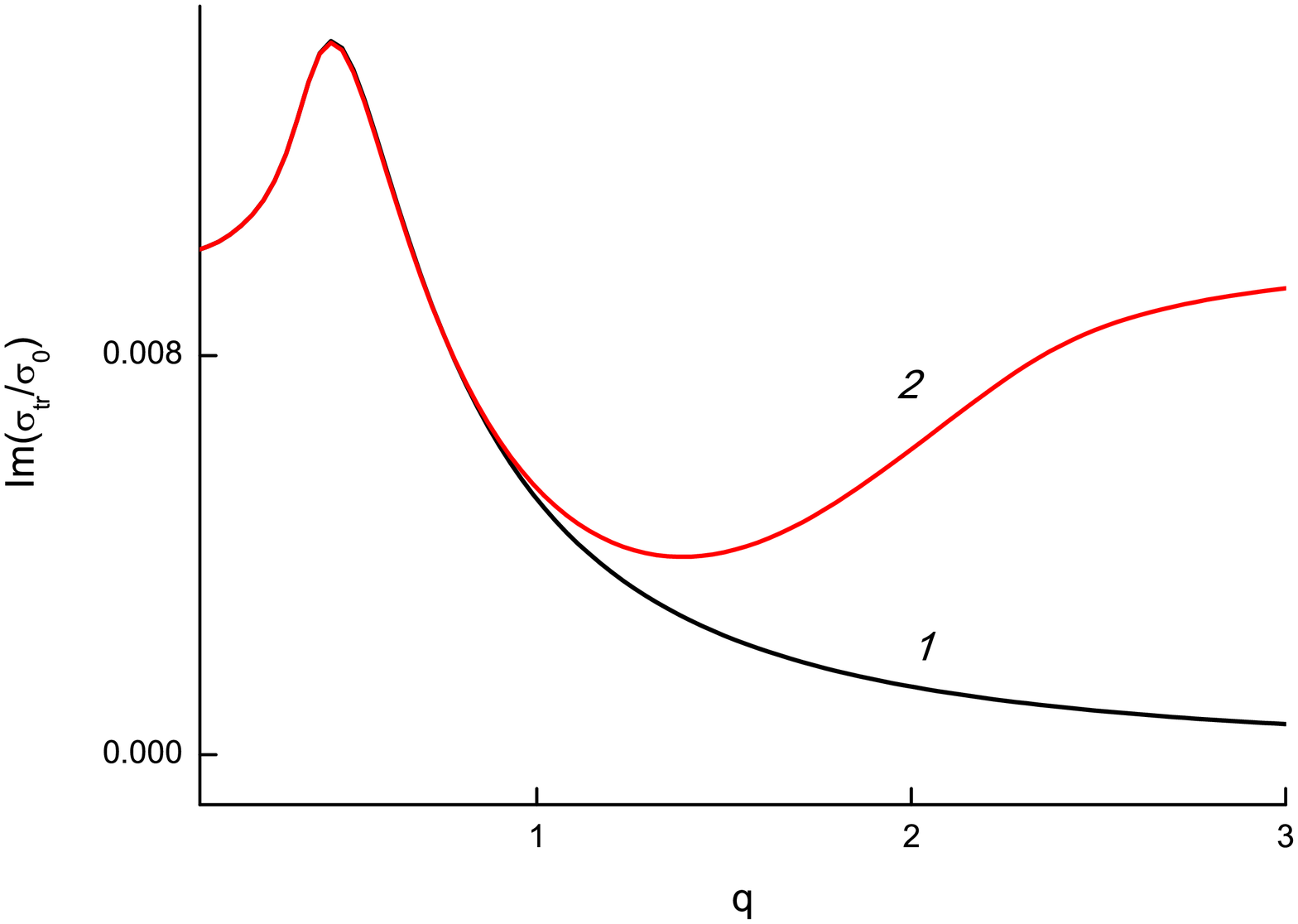}
\caption{Case: $x=1, y=0.01, \alpha=5.$
Dependence $\Re(\sigma_{tr}/\sigma_0)$ on quantity $q$.}
\end{center}
%\end{figure}
%\begin{figure}[h]
\begin{center}
\includegraphics[width=16cm, height=10cm]{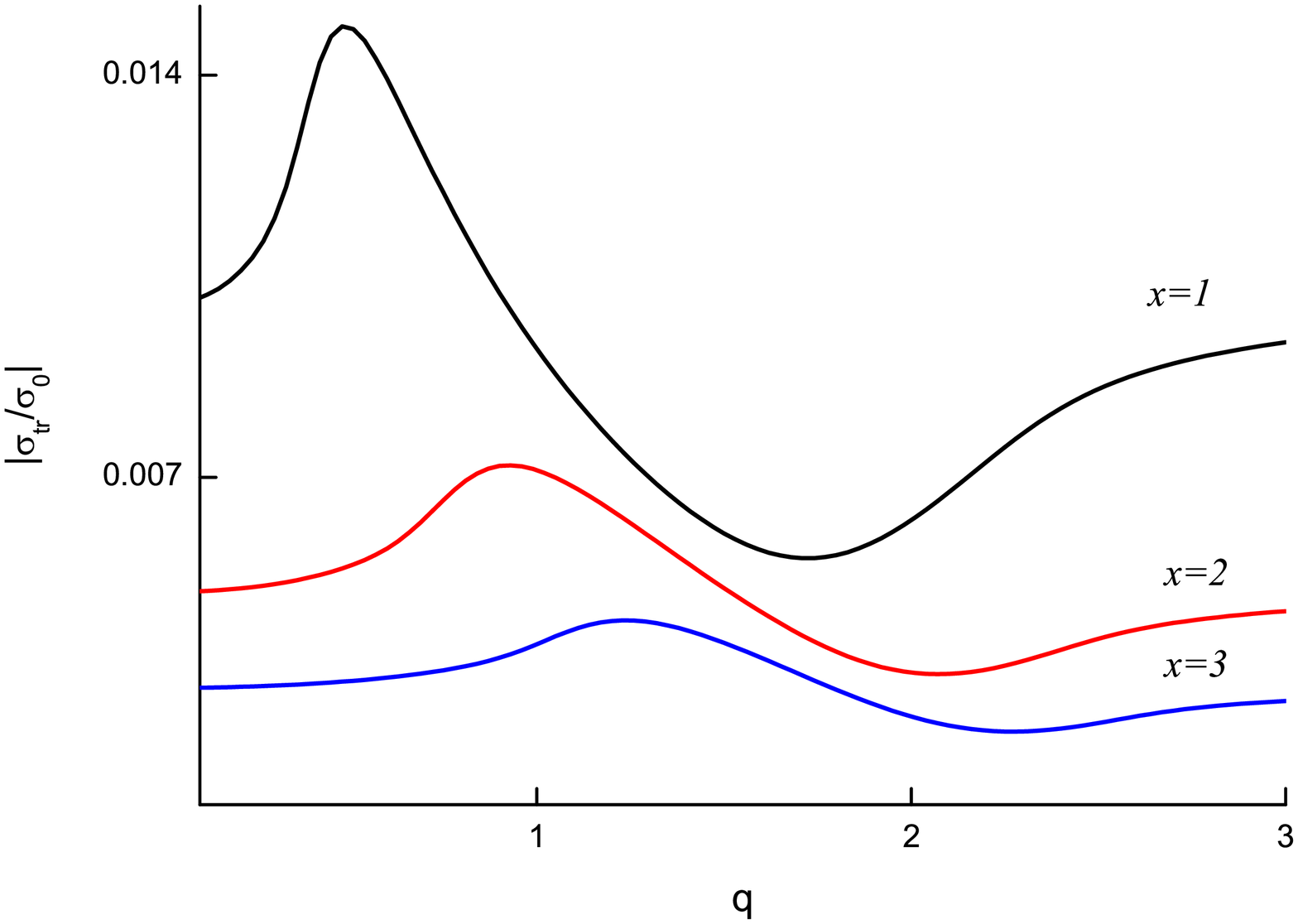}
\caption{Case: $x=1, y=0.01, \alpha=0.$
Dependen $|(\sigma_{tr}/\sigma_0)|$ on quantity $q$.}
\end{center}
\end{figure}


\clearpage

\begin{thebibliography}{99}

\renewcommand{\baselinestretch}{0.1}

\bibitem{Klim} {\it Klimontovich Y. and Silin V.P.} 
The Spectra of Systems of Interacting 
Particles // JETF (Journal Experimental Theoreticheskoi Fiziki), {\bf 23},
151 (1952).

\bibitem{Lin} {\it Lindhard J.} On the properties of a gas of
charged particles // 
Kongelige Danske Videnskabernes Selskab,
Matematisk--Fysiske Meddelelser. V. 28, \No 8 (1954), 1--57.

\bibitem{Roos2} {\it Von Roos O.} { Boltzmann --- Vlasov Equation
for a Quantum Plasma} // 
Phys. Rev. {\bf 119}. \No 4 (1960), 1174--1179.

\bibitem{Kliewer}{\it Kliewer K.L., Fuchs R.} Lindhard Dielectric Functions with a Finite Electron Lifetime // 
Phys. Rev. 1969. V. 181. \No 2. P. 552--558.

\bibitem{Mermin} {\it Mermin N. D.}{ Lindhard Dielectric Functions
in the Relaxation--Time Approximation} // Phys. Rev. B. 1970. V. 1, \No 5. P. 2362--2363.

\bibitem{Manf} {\it Manfredi G.} {How to model quantum plasmas} // 
ArXiv:quant-ph/0505004. 30 pp.

\bibitem{Anderson} {\it Anderson D., Hall B., Lisak M., and
Marklund M.}
  {Statistical effects in the multistream model for quantum plasmas}// 
Phys. Rev. E {\bf 65} (2002), 046417.

\bibitem{Andres}{\it De Andr\'{e}s P., Monreal R., and Flores F.}
{ Relaxation--time effects in the transverse
dielectric function and the
electromagnetic properties of metallic surfaces and small particles} /
Phys. Rev. {\bf B}. 1986. Vol. 34,\No 10, 7365--7366.

\bibitem{Shukla1} {\it  Shukla P.K. and Eliasson B.} 
Nonlinear aspects of quantum plasma physics 
//
Uspekhy Fiz. Nauk, {\bf 53}(1) 2010;
[V. 180. No. 1, 55-82 (2010) (in Russian)].

\bibitem{Shukla2}{\it Eliasson B. and Shukla P.K.}
Dispersion properties of electrostatic oscillations in quantum plasmas // 
arXiv:0911.4594v1 [physics.plasm-ph] 24 Nov 2009, 9 pp.

\bibitem{BGK} {\it Bhatnagar P.L., Gross E.P., and Krook M.}
{A model for collision processes in gases. I. 
Small amplitude processes in charged and neutral one-component systems} 
// Phys. Rev. {\bf 94} (1954), 511--525.

\bibitem{Opher}{\it Opher M., Morales G.J., Leboeuf J.N.}
Krook collisional models of the kinetic susceptibility of plasmas //
Phys. Rev. E. V. 66, 016407, 2002.

\bibitem{Gelder} {\it Gelder van, A.P.} {Quantum Corrections in the
Theory of the Anomalous Skin Effect} // Phys. Rev. 1969. Vol. 187. \No 3. P. 833--842.

\bibitem{Fuchs}{\it Fuchs R., Kliewer K.L.}
Surface plasmon in a
semi--infinite free--electron gas // Phys. Rev. B. 1971. V. 3. \No 7. P. 2270--2278.

\bibitem{Fuchs2}{\it Fuchs R., Kliewer K.L.}
 Optical properties of an
electron gas: further studies of a nonlocal description // 
Phys. Rev. 1969. V. 185. \No 3. P. 905--913.

\bibitem{Dressel}{\it Dressel M., Gr\"{u}ner G.} {Electrodynamics of Solids. Optical Properties of
Electrons in Matter} / Cambridge. Univ. Press. 2003. 487 p.

\bibitem{Wier} {\it Wierling A.} {Interpolation between local
field corrections and the Drude model by a generalized Mermin
approach} //
arXiv:0812.3835v1 [physics.plasm-ph] 19 Dec 2008.

\bibitem{Brod}{\it Brodin G., Marklund M., Manfredi G.}
{Quantum Plasma Effects in the Classical Regime} // 
Phys. Rev. Letters. {\bf 100}, (2008). P. 175001-1 -- 175001-4.

\bibitem{Manf2}{\it Manfredi G. and Haas F.}
{Self-consistent fluid model for a quantum electron gas} // 
Phys. Rev. B {\bf 64} (2001), 075316.

\bibitem{Wigner} {\it Wigner E.P.}
{On the quantum correction for thermodynamic equilibrium} // 
Phys. Rev. {\bf 40} (1932), 749--759.

\bibitem{Tatarskii} {\it Tatarskii V.I.}
{The Wigner representation of quantum mechanics} //
Uspekhy Fiz. Nauk. {\bf 26} (1983), 311--327;
[Usp. Fis. Nauk. {\bf 139} (1983), 587 (in Russian)].

\bibitem{Hillery}{\it Hillery M., O'Connell R.F., Scully M.O., and Wigner E.P.}
{Distribution functions in physics: Fundamentals} //
Phys. Rev. {\bf 106} (1984), 121--167.

\bibitem{Pains}{\it Pains D, Nozi\`{e}r P.} The theory of quantum liquids//
Benjamin, New York, 1969.
                       
\end{thebibliography}
\end{document}